\documentclass[12pt]{article} 
\usepackage{amsmath,epsfig,latexsym,amssymb,axodraw,color,cite}  
\def\ltsim{\lower3pt\hbox{$\, \buildrel < \over \sim \, $}}  
\def\gtsim{\lower3pt\hbox{$\, \buildrel > \over \sim \, $}}  
\def\be{\begin{equation}}  
\def\ee{\end{equation}}  
\def\ba{\begin{eqnarray}}  
\def\ea{\end{eqnarray}}  
\def\ga{\mathrel{\raise.3ex\hbox{$>$\kern-.75em\lower1ex\hbox{$\sim$}}}}  
\def\la{\mathrel{\raise.3ex\hbox{$<$\kern-.75em\lower1ex\hbox{$\sim$}}}}

\newcommand{\de}{\partial}  
\begin{document}

\baselineskip=16pt   
\begin{titlepage}  
\rightline{OUTP-01-41P}  
\rightline{hep-th/0107086}  
\rightline{July 2001} 
\begin{center}  
  
\vspace{0.5cm}  
  
\large {\bf Multigravity in six dimensions:\\ Generating bounces with
flat positive tension branes}  
  
\vspace*{5mm}  
\normalsize  
  
{\bf Ian I. Kogan\footnote{i.kogan@physics.ox.ac.uk}, Stavros  
Mouslopoulos\footnote{s.mouslopoulos@physics.ox.ac.uk},  Antonios  
Papazoglou\footnote{a.papazoglou@physics.ox.ac.uk}\\ and Graham G. Ross\footnote{g.ross@physics.ox.ac.uk}}  
  
\smallskip   
\medskip   
{\it Theoretical Physics, Department of Physics, Oxford University}\\  
{\it 1 Keble Road, Oxford, OX1 3NP,  UK}  
\smallskip

\vskip0.6in \end{center}  
   
\centerline{\large\bf Abstract}  
 
We present a generalization of the five dimensional multigravity
models to six dimensions. The key characteristic of these
constructions is that that we obtain solutions which do not have any
negative tension branes while at the same time the branes are kept
flat. This is due to the fact that in six dimensions the internal space 
is not trivial and its curvature allows bounce configurations with the 
above feature. These constructions give for the first time a theoretically and
phenomenologically viable realization of multigravity.

\vspace*{2mm}   

\end{titlepage}  
  
\section{Introduction}  
  
Over the past two years there has been considerable activity studying
brane configurations in which gravity is modified in large scales motivated by higher 
dimensional constructions \cite{Kogan:2000wc,Mouslopoulos:2000er,Gregory:2000jc,Kogan:2000cv,Kogan:2000xc,Dvali:2000hr}. The interesting feature of these scenarios 
is that modification of gravity at cosmological scales was
triggered by microphysics considerations which have to do with the
structure of the extra dimensional space of usually Planck scale
volume. These scenarios are in principle testable by the CMB power
spectrum \cite{Binetruy:2000xv} and gravitational lensing
\cite{Uzan:2000mz} observations.

The prototype multigravity scenario was the $''+-+''$ bigravity model \cite{Kogan:2000wc} where, apart
of the graviton zero mode, an anomalously light graviton Kaluza-Klein (KK) state is
present. This gave the possibility that gravity at intermediate scales 
($1{\rm{mm}}<r<10^{26}{\rm{cm}}$) is mediated by \textit{both} the massless graviton
\textit{and} the ultralight state, leading to the terminology
``bigravity''. At large scales were the Yukawa suppression of the massive 
state turns on, the gravitational constant decreases to a value which
for example in the symmetric configuration is  half of its value at intermediate scales. For asymmetric configurations one can have a
situation were the the massive graviton dominates and gravity switches 
off at large scales or, on the other hand,  where the massless mode 
dominates and gravity does not change at all.  Similar constructions of freely moving negative tension branes can also
give qualitatively different modifications of gravity at large
scales. The quasi-localized gravity model of Gregory, Rubakov and
Sibiryakov \cite{Gregory:2000jc} has no normalizable zero mode but gravity at intermediate
distances is generated by a ``resonance''-like
\cite{Csaki:2000pp,Dvali:2000rv} coupling of the low
part of the continuum of the KK spectrum. In this construction, the
gravitational law at large scales changes from four dimensional to
five dimensional. The change of the nature of the
gravitational law was observed also in the the crystalline universe
model \cite{Kogan:2000xc} where gravity at intermediate distances is generated by the
first low lying band of KK states and at large scales gives again the five
dimensional law of gravity.

So far our discussion has considered the gravitational force 
due to tensorial gravitational modes. However, in the gravity
sector there exists a moduli corresponding to the fluctuation(s) of the
negative tension brane(s), the radion, which changes the large scale
phenomenology of these models. In particular, the radion is a ghost
field \cite{Dvali:2000km,Pilo:2000et,Kogan:2001qx}, {\textit{i.e.}} it has wrong sign kinetic term, so at large scales it
will dominate in all the above constructions giving scalar antigravity \cite{Gregory:2000iu}  
instead of a reduction of the Newton's constant or a change of the nature
of the gravitational law. However, the very fact that there is a
physical ghost in the spectrum makes these models unacceptable. Note
that the essential  characteristic of
all these models is the bounce of the warp factor generated by the
presence of the moving negative tension branes. This association
between the bounce in the warp factor and presence of  negative tension branes 
is an unavoidable feature in flat (five dimensional) brane models and is linked to the
fact that the weaker energy condition is violated at the position of
the bounce \cite{Freedman:1999gp,Witten:2000zk}. As a result one must
have the unacceptable ghost field.

A way out of this difficulty is to consider $AdS_4$ branes instead of
flat branes (see \cite{Gorsky:2000rz} for a different possibility
involving an external four-form field). In this case the warp factor has a $''\cosh''$ form which naturally
generates a bounce without any need of floating negative tension
branes. Indeed, it is straightforward to replicate the bigravity   and
the crystalline model in this framework. The bigravity model is converted to a
$''++''$ model \cite{Kogan:2001vb} and the crystal model to an infinite array of $''+''$
branes. The quasi-localized model on the other hand cannot be
reproduced because the presence of the cosmological constant on the
branes prevents the warp factor from being asymptotically constant. By
sending the second brane in the $''++''$ model to infinity, one obtains 
the locally localized model of \cite{Karch:2001ct,Karch:2001cw} where gravity is
mediated only by the ultralight state since the zero mode becomes
non-normalizable. These models, even though they solve
the theoretical difficulty of the ghost radion(s) in the flat brane
models, face phenomenological difficulties. The presence of a remnant 
negative cosmological constant is at odds with observations, and
furthermore it turns out that all large scale modifications of
gravity predicted by these models, are hidden  behind the 
$AdS$ horizon.

It is interesting to note at this point how these models confront the
relativistic predictions of four dimensional Einsteinian gravity. It
has been known for a long time that the structure of the massive graviton
propagator is different form the massless one in flat space, something 
that is known as the van Dam - Veltman - Zakharov discontinuity \cite{vanDam:1970vg,Zakharov}. This 
is due to the extra polarization states of the massive graviton and
in particular the scalar-like mode that does not decouple in the limit 
 of vanishing mass. As a result, gravity generated by a massive
graviton case predicts a $25\%$ discrepancy in
the bending of light observation if one insists in the validity of 
Newton's law for static bodies or vice versa. This apparently rules out all
models of gravity based on a massive graviton \cite{Dvali:2000rv}
since the observed bending of light by the Sun
agrees with the General Relativistic result to an accuracy of $1\%$
(see also \cite{Tekin:2001xj} for a  discussion on the relation of this observation with physics arising from higher curvature terms
in four dimensions). In
the models of the flat branes, the ghost radion(s) at intermediate
distances cancel the troublesome additional polarization state(s) of
the massive graviton(s),
restoring the phenomenologically favoured tensorial structure of the
graviton propagator \cite{Pilo:2000et,Gregory:2000iu,Csaki:2000ei}. In the models with $AdS$ branes it was shown that
the extra polarization state is practically decoupled due to an
unusual feature of the graviton propagator structure in $AdS$. In more 
detail, it was shown \cite{Higuchi:1987py,Higuchi:1989gz,Kogan:2001uy,Porrati:2001cp}   that the van Dam - Veltman - Zakharov no-go
theorem can be evaded at tree level in $(A)dS$ in the case where $m/H \to 0$, where
$m$ is the mass of the graviton and  $H$ the ``Hubble'' of the $AdS$
space, a precise realization of which was the $''++''$ brane
model. Thus, in the $AdS$ branes no phenomenological constraint from
the bending of light experiment is applicable provided this condition holds. The discontinuity in
the pure four dimensional theory with a massive graviton reappears at the quantum level
\cite{Dilkes:2001av,Duff:2001zz} but one should keep in mind that
purely four dimensional theory of massive gravity is not well
defined \cite{Boulware:1972my}. If instead one starts from the higher
dimensional theory in which the mass terms appear dynamically and if one takes
into account the whole tower of the KK states, this quantum
discontinuity is expected to be absent, as pointed out in
\cite{Kogan:2001qx}. Let us also note that in a recent paper
\cite{Karch:2001jb} it was suggested that the extra polarizations of
the massive state in the one brane model can be gauged away and thus
the discontinuity is absent.

More generally, the argument of   van Dam - Veltman - Zakharov that no mass
for the graviton is allowed in a flat background is not entirely
robust. A long time ago Vainshtein \cite{Vain} showed that if one considers the
bending of light, the lowest order approximation
is not valid because the calculation does not take into account
the characteristic mass scale of the problem. A more detailed non-perturbative analysis presented in the recent
paper  \cite{Deffayet:2001uk} supports the original Vainshtein argument that there is continuity 
in the gravitational potential as the mass of the graviton goes to zero for distances smaller that a critical
one which, in the case of the Sun, is of order of the size of the solar
system. The linear approximation is valid for distances greater than 
this critical one, where the discontinuity reappears. Thus the
observation of light bending by the Sun does not rule out the
massive graviton proposal.

In the light of this encouraging result it seems appropriate to look for 
multigravity  models of flat branes without ghost fields. As we have
discussed, in five dimensions it is
impossible to have flat brane multigravity models without negative
tension branes. For this reason, we will consider models in six dimensions. The
literature on six dimensional constructions is already quite rich 
\cite{Chodos:1999zt,Cohen:1999ia,Gregory:2000gv,Chacko:2000eb,Arkani-Hamed:2000dz,Chen:2000at,Olasagasti:2000gx,Gherghetta:2000qi,Chodos:2000tf,Gherghetta:2000jf,Ponton:2001gi,Moon:2001hn,Charmousis:2001hg,Corradini:2001su,Collins:2001ni,Kim:2001rm,Kanti:2001vb,Hayakawa:2001ke,Leblond:2001xr}.
In this paper we explicitly show that the difficulties of the five
dimensional models can be evaded in six dimensions. It is possible to construct
multigravity models
with only flat positive tension branes. The branes which localize gravity in
this setup are four-branes, but one of their dimensions is compact,
unwarped and of Planck length. Thus, in the low energy limit the
spacetime on the brane appears three dimensional. In order that these
constructions are realized it is crucial that the tensions and/or the bulk cosmological constant are inhomogeneous. The five dimensional constructions in this
setup come into two types. One of them involves conical singularities 
at finite distance from the four branes. These conical singularities
support three-branes\footnote{We would like to thank Panagiota Kanti and
Luigi Pilo for bringing this issue to our attention.} which can be of
positive tension if one has an
angle deficit, of zero tension if one has no angle deficit and of
negative tension if one has an angle excess. The other type of the
multigravity models has no conical singularity at all. It is crucial
to note that there is no five dimensional effective theory for these
constructions, otherwise one would get all the problems faced in the
five dimensional constructions. The low energy effective theory is
directly four dimensional.

The organization of the paper is as following. In section two we present
the cylindrically symmetric single four-brane warped models and show the cases 
where gravity is localized on the four-branes. In section three we paste two 
of the single four-brane solutions and obtain the double four-brane bigravity
models. In section four we generalize the quasi-localized and
crystalline constructions in six dimensions. Finally, we 
summarize the six dimensional cases and conclude.

\section{Single brane models in six dimensions}

At first we will discuss the single brane solutions in six dimensions to
get an insight on the more complicated multigravity
configurations. In the following subsection  we will review the
minimal model where the bulk cosmological constant is
homogeneous and then a more generalized model with arbitrary
cosmological constant components along the four dimensional and the
extra dimensional directions. In the following we will implicitly
assume orbifolding around the four-brane.

\subsection{The minimal single brane model}

The simplest single brane model consists of a four-brane embedded
in six dimensional $AdS$ space \cite{Gherghetta:2000qi}. One of the longitudinal dimensions of
the four brane is compactified to a Planck length radius $R$ while the dimensions
transverse to the four-brane compact dimension are infinite.  The most general spherically symmetric ansatz that one can write down
in six dimensions and which preserves four dimensional Poincar\'e invariance  is\footnote{Note
that in this paper we use different metric signature and different
definition of the fundamental scale from \cite{Gherghetta:2000qi}.}:
\begin{equation}
ds^2=\sigma(\rho)\eta_{\mu \nu}dx^{\mu}dx^{\nu}+d\rho^2+\gamma(\rho)d\theta^2
\end{equation}
where $\theta$ is the compactified dimension with range $[0,2\pi]$ and
$\rho$ is the infinite radial dimension.

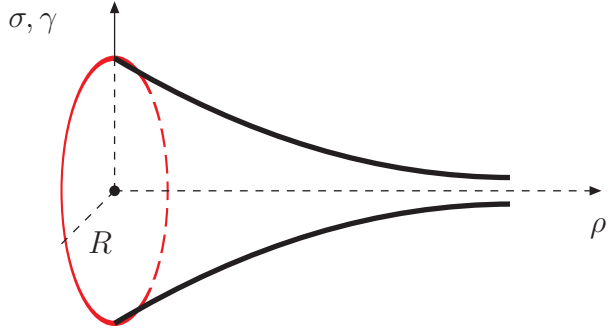
\begin{figure}[t]
\vskip10mm
\begin{center}
\begin{picture}(200,100)(0,50)

\SetWidth{2}
\SetColor{Red}
\Oval(0,100)(50,20)(0)
\SetColor{Black}
\Curve{(0,150)(150,105)(300,150)}
\Curve{(0,50)(150,95)(300,50)}
\Vertex(0,100){2}
\Text(-40,160)[lb]{$\sigma,\gamma$}

\CBox(150,50)(300,150){White}{White}
\SetWidth{.5}
\DashLine(0,100)(180,100){3}
\LongArrow(180,100)(182,100)

\LongArrow(0,150)(0,170)
\DashLine(0,100)(0,150){3}
\DashLine(0,100)(-20,80){3}

\SetWidth{3}
\SetColor{White}
\Line(10,135)(20,135)
\Line(10,125)(20,125)
\Line(10,115)(20,115)
\Line(10,105)(25,105)
\Line(10,95)(25,95)
\Line(10,85)(20,85)
\Line(10,75)(20,75)
\Line(10,65)(20,65)

\SetWidth{2}
\SetColor{Black}
\Text(180,90)[lt]{$\rho$}
\Text(-10,85)[lt]{$R$}

\end{picture}
\end{center}

\caption{The minimal single four-brane model warp factors $\sigma(\rho)$
and $\gamma(\rho)$.}
\label{1min}
\end{figure}

It is straightforward to solve the Einstein equations for the bulk
energy momentum tensor $T^{(B)~N}_{~~~M}=-\Lambda \delta_M^N$ with the brane contribution:
\begin{equation}
T^{(br)~N}_{~~~M}=-\delta(\rho)\left(\begin{array}{ccc}V
~\delta_{\mu}^{\nu}&~&~\\~&0&~\\~&~&V \end{array}\right)
\end{equation}

From the form of the Einstein equations given in the Appendix we see
that the solution for the two warp factors is: 
\begin{equation}
\sigma(\rho)=e^{-k \rho} ~~~,~~~ \gamma(\rho)=R^2 e^{-k \rho}
\end{equation}
with $k^2=-{\Lambda \over 10M^4}$, where $M$ is the six dimensional
fundamental scale and  the arbitrary integration constant $R$ is just the radius of
the four-brane (see fig.\ref{1min}). The Einstein equations require the usual fine tuning
between the bulk cosmological constant and the tension of the four brane:
\begin{equation}
V=-{8\Lambda \over 5k}
\end{equation}
Let us note at this point that in \cite{Gherghetta:2000qi} this fine tuning was absent
because a smooth local defect was considered instead of a four-brane. In this case, the fine tuning emerges between the different
components of the defect energy momentum tensor. The physics of the
four-brane idealization and the one of the defect model is the same.

The four dimensional KK decomposition can be carried out as
usual (see \cite{Csaki:2000fc} for a detailed fluctuation analysis) by considering the following graviton perturbations:
\begin{equation}
ds^2=\sigma(\rho)\left[\eta_{\mu \nu}+h_{\mu \nu}(\rho,\theta,x)\right]dx^{\mu}dx^{\nu}+d\rho^2+\gamma(\rho)d\theta^2
\end{equation}

Here, as well as throughout this paper, we have ignored the modulus
associated to the radius $R$ of the four brane. This will be a
massless scalar and in order not to give rise to an additional long range
force it should be given mass through some stabilization mechanism.

We expand the graviton perturbations in a complete set of radial
eigenfunctions and Fourier angular modes:  
\begin{equation}
h_{\mu \nu}(\rho,\theta,x)=\sum_{n,l} \phi_{(n,l)}(\rho)e^{il\theta} h^{(n,l)}_{\mu \nu}(x)
\end{equation}

The differential equation for the radial wavefunctions $\phi$ is:
\begin{equation}
\phi''-{5 \over 2}k\phi'+\left(m^2-{l^2 \over R^2}\right)e^{k\rho}\phi=0
\end{equation}
with normalization  $\int_0^\infty d\rho \sigma \sqrt{\gamma} \phi_m^*
\phi_n=\delta_{mn}$. We can convert this equation to a two dimensional 
Schr\"{o}dinger-like equation by the following redefinitions:
\begin{equation}
z={2 \over k}\left(e^{{k \over 2}\rho}-1\right)~~~,~~~\hat{\Psi}=\sigma^{3/4}\phi
\end{equation}
so that 
\begin{equation}
-{1 \over 2 \sqrt{\gamma}} \de_z\left(\sqrt{\gamma}\de_z\hat{\Psi}\right)+V_{eff}\hat{\Psi}={m^2 \over 2}\hat{\Psi}~~~,~~~V_{eff}(z)=\frac{21k^2}{32\left({kz \over
2}+1\right)^2}+{l^2 \over 2R^2}-{3k \over 4}\delta(z)
\end{equation}

\begin{figure}[t]
\vskip10mm
\begin{center}
\begin{picture}(200,100)(0,50)

\SetWidth{2}
\SetColor{Black}
\Curve{(-50,110)(150,5)(350,110)}
\SetColor{Red}
\Curve{(-50,130)(150,25)(350,130)}
\SetColor{Green}
\Curve{(-50,150)(150,45)(350,150)}
\CBox(150,5)(350,150){White}{White}
\SetColor{Black}
\Curve{(149,5)(200,3)(230,2)}
\SetColor{Red}
\Curve{(149,25)(200,23)(230,22)}
\SetColor{Green}
\Curve{(149,45)(200,43)(230,42)}
\SetColor{Black}
\Vertex(-50,0){2}

\SetWidth{.5}
\LongArrow(-50,0)(250,0)
\LongArrow(-50,0)(-50,170)

\SetWidth{2}
\SetColor{Black}
\Text(250,20)[lt]{$z$}
\Text(-40,160)[lb]{$V_{eff}$}
\Text(70,100)[lt]{$l=0$}
\Text(70,120)[lt]{${\Red{l=1}}$}
\Text(70,140)[lt]{${\Green{l=2}}$}
\end{picture}
\end{center}
\vskip15mm
\caption{The form of the effective potential $V_{eff}$ for different angular
quantum numbers $l$ for the minimal single four-brane model. The $\delta$-function in the origin is omitted.}
\label{pot1min}
\end{figure}
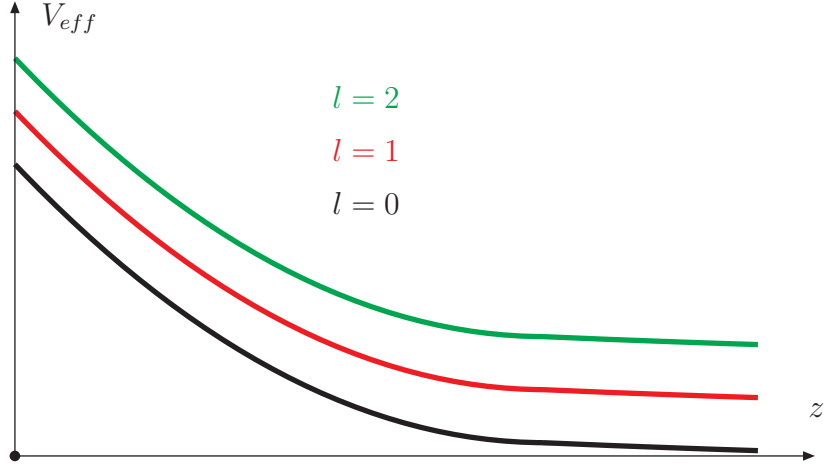

From the form of the potential (see fig.\ref{pot1min}) we can easily deduce that the angular
excitations spectrum will consist of continuum starting from a gap of
the order ${l^2 \over R^2}$ and thus can be safely ignored. For the s-wave ($l=0$) there is a normalizable zero mode which is a
constant in the $\rho$ coordinate, \textit{i.e.} $\phi_0=const.$ The
KK tower for the s-waves will again form a continuum but this time
gapless with wavefunctions given by \cite{Gherghetta:2000qi}:
\begin{equation}
\phi_m=N_m e^{{5 \over 4}k \rho}\left[J_{3/2}\left({2m \over k}\right)Y_{5/2}\left({2m \over k}e^{{k \over 2} \rho}\right)-Y_{3/2}\left({2m \over k}\right)J_{5/2}\left({2m \over k}e^{{k \over 2} \rho}\right)\right]
\end{equation}

The correction to the Newton's law due to the s-modes can be easily calculated and one
finds that it is more suppressed than the five dimensional
Randall-Sundrum   case \cite{Csaki:2000fc,Gherghetta:2000qi,Randall:1999vf}. In six dimensions the 
correction reads:
\begin{equation}
\Delta V=-{1 \over {\mathcal{O}}(M_{Pl}^5)}{1 \over r^4}
\end{equation}

\subsection{The generalized single brane model}

We can now try to find more general single four-brane solutions by relaxing
the requirement of homogeneity of the bulk and brane energy momentum
tensors. In contrast to
the five dimensional case,  this is possible in six dimensions because there are two independent functions
in the metric (see Appendix). We will consider the following inhomogeneous  bulk energy-momentum tensor: 
\begin{equation}
T^{(B)~N}_{~~~M}=-\left(\begin{array}{ccc}\Lambda_0
~\delta_{\mu}^{\nu}&~&~\\~&\Lambda_{\rho}&~\\~&~&\Lambda_{\theta}\end{array}\right)
\label{bulkT}
\end{equation}
and allow for an inhomogeneous brane tension of the form: 
\begin{equation}
T^{(br)~N}_{~~~M}=-\delta(\rho)\left(\begin{array}{ccc}V_0~
\delta_{\mu}^{\nu}&~&~\\~&0&~\\~&~&V_{\theta}\end{array}\right)
\label{braneT}
\end{equation}
This  inhomogeneity can be due to different contributions to the
Casimir energy in the different directions
\cite{Kogan:1983fp,Candelas:1984ae}  (see also \cite{Appelquist:1987nr}) or due to a background
three-form gauge field with non-zero field strength \cite{Chen:2000at}.

\begin{figure}[b!]
\vskip10mm
\begin{center}
\begin{picture}(200,100)(0,50)

\SetWidth{2}
\SetColor{Red}
\Oval(0,100)(50,20)(0)
\SetColor{Black}
\Curve{(0,150)(150,105)(300,150)}
\Curve{(0,50)(150,95)(300,50)}
\Vertex(0,100){2}
\Text(-70,160)[lb]{$\sigma,\left.\gamma\right|_{\alpha>6/5}$}

\CBox(150,50)(300,150){White}{White}
\SetWidth{.5}
\DashLine(0,100)(180,100){3}
\LongArrow(180,100)(182,100)

\LongArrow(0,150)(0,170)
\DashLine(0,100)(0,150){3}
\DashLine(0,100)(-20,80){3}

\SetWidth{3}
\SetColor{White}
\Line(10,135)(20,135)
\Line(10,125)(20,125)
\Line(10,115)(20,115)
\Line(10,105)(25,105)
\Line(10,95)(25,95)
\Line(10,85)(20,85)
\Line(10,75)(20,75)
\Line(10,65)(20,65)

\SetWidth{2}
\SetColor{Black}
\Text(180,90)[lt]{$\rho$}
\Text(-10,85)[lt]{$R$}

\end{picture}
\end{center}

\vskip5mm

\vskip5mm
\begin{center}
\begin{picture}(200,100)(0,50)

\SetWidth{2}
\SetColor{Red}
\Oval(-110,100)(50,20)(0)
\SetColor{Black}
\Line(-110,150)(40,150)
\Line(-110,50)(40,50)
\Vertex(-110,100){2}
\Text(-160,160)[lb]{$\left.\gamma\right|_{\alpha=6/5}$}

\SetWidth{.5}
\DashLine(-110,100)(70,100){3}
\LongArrow(70,100)(72,100)

\LongArrow(-110,150)(-110,170)
\DashLine(-110,100)(-110,150){3}
\DashLine(-110,100)(-130,80){3}

\SetWidth{3}
\SetColor{White}
\Line(-100,135)(-90,135)
\Line(-100,125)(-90,125)
\Line(-100,115)(-90,115)
\Line(-100,105)(-85,105)
\Line(-100,95)(-85,95)
\Line(-100,85)(-90,85)
\Line(-100,75)(-90,75)
\Line(-100,65)(-90,65)

\SetWidth{2}
\SetColor{Black}
\Text(70,90)[lt]{$\rho$}
\Text(-120,85)[lt]{$R$}

\SetWidth{2}
\SetColor{Red}
\Oval(110,100)(50,20)(0)
\SetColor{Black}
\Curve{(110,150)(200,160)(260,180)}
\Curve{(110,50)(200,40)(260,20)}
\Vertex(110,100){2}
\Text(60,160)[lb]{$\left.\gamma\right|_{\alpha<6/5}$}

\SetWidth{.5}
\DashLine(110,100)(290,100){3}
\LongArrow(290,100)(292,100)

\LongArrow(110,150)(110,170)
\DashLine(110,100)(110,150){3}
\DashLine(110,100)(90,80){3}

\SetWidth{3}
\SetColor{White}
\Line(120,135)(130,135)
\Line(120,125)(130,125)
\Line(120,115)(130,115)
\Line(120,105)(135,105)
\Line(120,95)(135,95)
\Line(120,85)(130,85)
\Line(120,75)(130,75)
\Line(120,65)(130,65)

\SetWidth{2}
\SetColor{Black}
\Text(290,90)[lt]{$\rho$}
\Text(100,85)[lt]{$R$}

\end{picture}
\end{center}

\vskip5mm
\caption{The generalized single four-brane warp factors $\sigma(\rho)$
and $\gamma(\rho)$ for different values of the parameter $\alpha$.}
\label{1gen}
\end{figure}
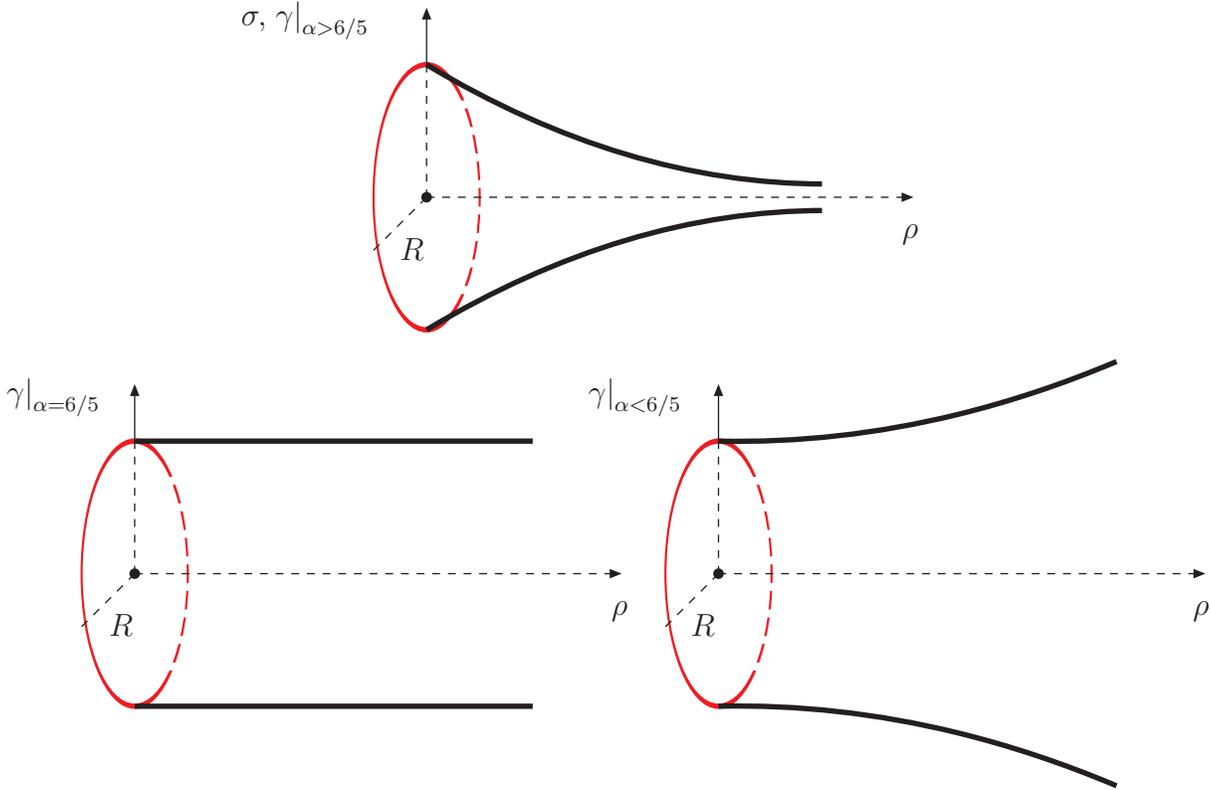

If we define the parameter $\alpha={2\Lambda_{\rho} \over
\Lambda_{\theta}}$, then the one four-brane solutions for the warp factors are:
\begin{equation}
\sigma(\rho)=e^{-k \rho} ~~~,~~~\gamma(\rho)=R^2 e^{-{k \over 4}(5\alpha-6) \rho}
\end{equation}
with $k^2=-{\Lambda_{\theta} \over 10M^4}$, where again $R$ is the
radius of the four-brane (see fig.\ref{1gen}). The flatness of the
brane is achieved by the following fine tunings of the the parameters
of the bulk and brane energy momentum tensors:
\begin{equation}
V_{\theta}=-{8\Lambda_{\theta} \over 5k} ~~~,~~~V_0={5\alpha+6 \over
16}V_{\theta}~~~,~~~\Lambda_0={5\alpha^2+12 \over 32}\Lambda_{\theta}
\end{equation}

\begin{figure}[t]
\vskip10mm
\begin{center}
\begin{picture}(200,100)(0,50)

\SetWidth{2}
\SetColor{Black}
\Curve{(-50,110)(150,5)(350,110)}
\CBox(150,5)(350,150){White}{White}
\SetColor{Black}
\Curve{(149,5)(200,3)(230,2)}
\SetColor{Red}
\Curve{(-50,130)(110,35)(270,130)}
\SetColor{Green}
\Curve{(-50,150)(90,60)(230,180)}
\CBox(230,70)(270,190){White}{White}
\SetColor{Black}
\Vertex(-50,0){2}

\SetWidth{.5}
\LongArrow(-50,0)(250,0)
\LongArrow(-50,0)(-50,170)

\SetWidth{2}
\SetColor{Black}
\Text(250,20)[lt]{$z$}
\Text(-40,160)[lb]{$V_{eff}$}
\Text(70,100)[lt]{$l=0$}
\Text(70,120)[lt]{${\Red{l=1}}$}
\Text(70,140)[lt]{${\Green{l=2}}$}

\end{picture}
\end{center}

\vskip15mm

\caption{The form of the effective potential $V_{eff}$ for $\alpha>2$ for different angular
quantum numbers $l$. The $\delta$-function in the origin is omitted.}
\label{pot1gen1}
\end{figure}

\begin{figure}[b!]
\vskip10mm
\begin{center}
\begin{picture}(200,100)(0,50)

\SetWidth{2}
\SetColor{Black}
\Curve{(-50,110)(150,5)(350,110)}
\SetColor{Red}
\Curve{(-50,130)(150,5)(350,130)}
\SetColor{Green}
\Curve{(-50,150)(150,5)(350,150)}
\CBox(150,5)(350,150){White}{White}
\SetColor{Black}
\Curve{(149,5)(200,3)(230,2)}
\SetColor{Red}
\Curve{(149,5)(200,3)(230,2)}
\SetColor{Green}
\Curve{(149,5)(200,3)(230,2)}
\SetColor{Black}
\Vertex(-50,0){2}

\SetWidth{.5}
\LongArrow(-50,0)(250,0)
\LongArrow(-50,0)(-50,170)

\SetWidth{2}
\SetColor{Black}
\Text(250,20)[lt]{$z$}
\Text(-40,160)[lb]{$V_{eff}$}
\Text(70,100)[lt]{$l=0$}
\Text(70,120)[lt]{${\Red{l=1}}$}
\Text(70,140)[lt]{${\Green{l=2}}$}

\end{picture}
\end{center}

\vskip15mm
\caption{The form of the effective potential $V_{eff}$ for $\alpha<2$ for different angular
quantum numbers $l$. The $\delta$-function in the origin is omitted.}
\label{pot1gen2}
\end{figure}

Let us note that the two components of the brane tension have 
the same sign as long as $\alpha>-{6 \over 5}$. For $\alpha>{6 \over
5}$ the internal space is shrinking as in \cite{Gherghetta:2000qi},  for $\alpha<{6 \over 5}$ the internal space is growing and
for $\alpha={6 \over 5}$ the internal space is a cylinder (see fig.\ref{1gen}).

Proceeding with the usual KK decomposition (see \cite{Csaki:2000fc}) we find the following
equation for the  radial wavefunction $\phi$:
\begin{equation}
\phi''-{5\alpha +10 \over 8}k\phi'+\left(m^2-{l^2 \over R^2}e^{{5
\over 4}(\alpha-2)k \rho}\right)e^{k\rho}\phi=0
\end{equation}
with normalization  $\int_0^\infty d\rho \sigma \sqrt{\gamma} \phi_m
\phi_n=\delta_{mn}$. With the wavefunction redefinition and the
coordinate change discussed in the previous subsection, we obtain the
following effective potential for the two dimensional Schr\"{o}dinger-like equation:
\begin{equation}
V_{eff}(z)=\frac{3 (5\alpha+4)k^2}{64\left({kz \over
2}+1\right)^2}+{l^2 \over 2R^2}\left({kz \over 2}+1\right)^{{5 \over 2}(\alpha-2)}-{3k \over 4}\delta(z)
\end{equation}

For the s-wave there is a normalizable zero mode when $\alpha>-{2
\over 5}$ which is a
constant in the $\rho$ coordinate, \textit{i.e.} $\phi_0=const.$ The
KK tower for the s-waves will again form a continuum and their
wavefunctions are given by:
\begin{equation}
\phi_m=N_m e^{{\nu \over 2}k \rho}\left[J_{\nu-1}\left({2m \over k}\right)Y_{\nu}\left({2m \over k}e^{{k \over 2} \rho}\right)-Y_{\nu-1}\left({2m \over k}\right)J_{\nu}\left({2m \over k}e^{{k \over 2} \rho}\right)\right]
\end{equation}
where $\nu={5 \over 8}(\alpha+2)$. The correction to the Newton's law due to the s-modes can be easily calculated and one
finds that it is negligible as long as $\alpha \gtsim -{2 \over 5}$ \cite{Csaki:2000fc}:

\begin{equation}
\Delta V=-{1 \over {\mathcal{O}}(M_{Pl}^{2\nu})}{1 \over r^{2\nu-1}}
\end{equation}

As far as the angular excitations $l\neq0$ are concerned, in the case where
$\alpha>2$ the potential diverges at infinity (see fig. \ref{pot1gen1}), so the spectrum will be 
discrete starting from a scale bigger than ${l^2 \over R^2}$ and thus
can be safely ignored. On the other hand, the potential for the case
$\alpha<2$ asymptotically vanishes (see fig. \ref{pot1gen2}) and so the spectrum will be gapless
continuum. One should examine the contribution of the angular excitations to
the Newtonian potential in this case.

\section{Bigravity in six dimensions}

The flat single four-brane models considered in the previous sections have the
characteristic that gravity is localized on the brane in the same way
as in the five dimensional analogue \cite{Randall:1999vf}. In this section we will
show how we can construct realistic multi-localization scenarios for
gravity by sewing two single four-brane solutions together. We will
present two explicit examples of compact double four-brane models where bigravity is realized. Firstly, we will consider a
double four-brane bigravity model which in general
contains an additional three-brane associated with the existence of a conical
singularity. This can be done by allowing for an inhomogeneous four-brane 
tension. In the following subsection we demonstrate how one can avoid the conical singularity by also allowing for inhomogeneous bulk
cosmological constant. In the following we will consider the
 orbifold geometry $S^1/Z_2$ in which the two four-branes lie on the
fixed points. 

\subsection{The conifold model}

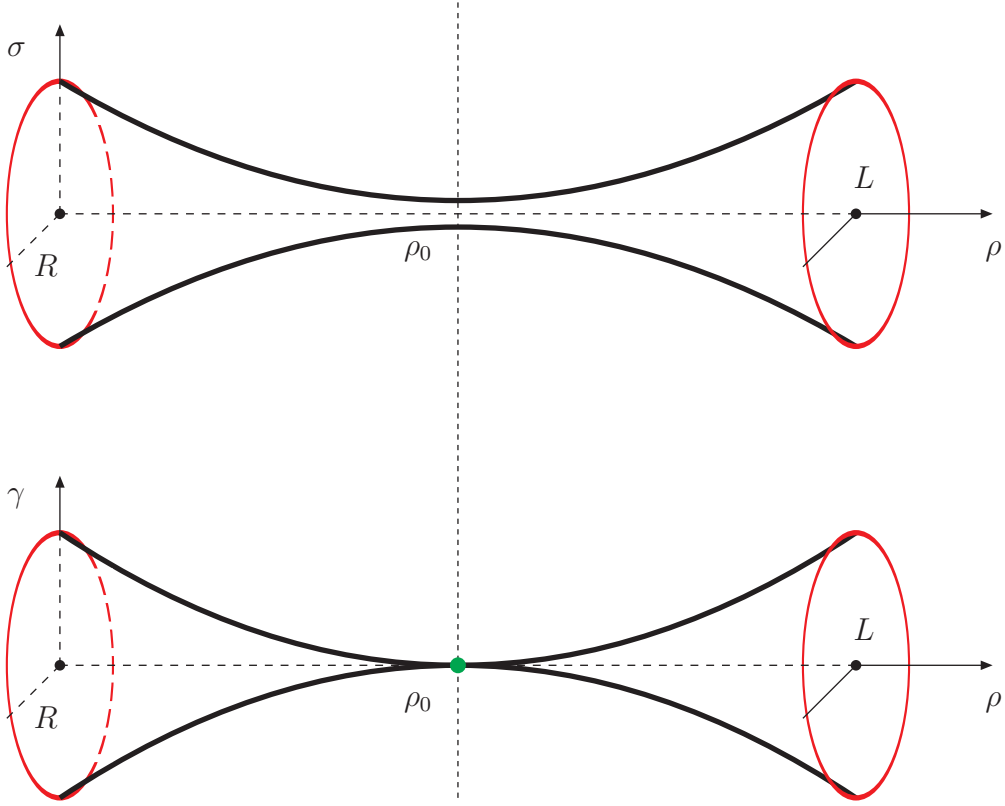
\begin{figure}[t]
\vskip10mm
\begin{center}
\begin{picture}(200,100)(0,50)

\SetWidth{2}
\SetColor{Red}
\Oval(-50,100)(50,20)(0)
\SetColor{Black}
\Curve{(-50,150)(100,105)(250,150)}
\Curve{(-50,50)(100,95)(250,50)}
\Vertex(-50,100){2}
\Vertex(250,100){2}

\SetColor{Red}
\Oval(250,100)(50,20)(0)

\SetColor{Black}
\SetWidth{.5}
\LongArrow(-50,150)(-50,170)
\DashLine(-50,100)(-50,150){3}
\DashLine(-50,100)(-70,80){3}
\Line(250,100)(230,80)

\SetColor{Black}
\SetWidth{.5}
\DashLine(-50,100)(250,100){3}
\LongArrow(250,100)(300,100)

\SetWidth{3}
\SetColor{White}
\Line(-40,135)(-30,135)
\Line(-40,125)(-30,125)
\Line(-40,115)(-30,115)
\Line(-40,105)(-25,105)
\Line(-40,95)(-25,95)
\Line(-40,85)(-30,85)
\Line(-40,75)(-30,75)
\Line(-40,65)(-30,65)

\SetWidth{2}
\SetColor{Black}
\Text(300,90)[lt]{$\rho$}
\Text(80,90)[lt]{$\rho_0$}
\Text(250,110)[lb]{$L$}
\Text(-60,85)[lt]{$R$}
\Text(-70,160)[lb]{$\sigma$}

\end{picture}
\end{center}

\vskip5mm

\vskip15mm
\begin{center}
\begin{picture}(200,100)(0,50)

\SetWidth{2}
\SetColor{Red}
\Oval(-50,100)(50,20)(0)
\SetColor{Black}
\Curve{(-50,150)(100,100)(250,150)}
\Curve{(-50,50)(100,100)(250,50)}
\Vertex(-50,100){2}
\Vertex(250,100){2}

\SetColor{Red}
\Oval(250,100)(50,20)(0)

\SetColor{Black}
\SetWidth{.5}
\LongArrow(-50,150)(-50,170)
\DashLine(-50,100)(-50,150){3}
\DashLine(-50,100)(-70,80){3}
\Line(250,100)(230,80)

\SetColor{Black}
\SetWidth{.5}
\DashLine(-50,100)(250,100){3}
\LongArrow(250,100)(300,100)
\SetColor{Black}
\SetWidth{.5}
\DashLine(100,50)(100,350){2}

\SetWidth{2}
\SetColor{Green}
\Vertex(100,100){3}

\SetWidth{3}
\SetColor{White}
\Line(-40,135)(-30,135)
\Line(-40,125)(-30,125)
\Line(-40,115)(-30,115)
\Line(-40,105)(-25,105)
\Line(-40,95)(-25,95)
\Line(-40,85)(-30,85)
\Line(-40,75)(-30,75)
\Line(-40,65)(-30,65)

\SetWidth{2}
\SetColor{Black}
\Text(300,90)[lt]{$\rho$}
\Text(80,90)[lt]{$\rho_0$}
\Text(250,110)[lb]{$L$}
\Text(-60,85)[lt]{$R$}
\Text(-70,160)[lb]{$\gamma$}

\end{picture}
\end{center}

\caption{The conifold bigravity model warp factors $\sigma(\rho)$ and
$\gamma(\rho)$ for the symmetric configuration where $L=2\rho_0$. Both 
the $\sigma(\rho)$ and $\gamma(\rho)$ have a bounce at
$\rho_0$. However,  $\gamma(\rho)$ vanishes at this point. This point in general corresponds to a conical singularity.}
\label{bigrcon}
\end{figure}

We are interested in a double four-brane generalization of the minimal
single four-brane model. In order to achieve this,  we still assume a
homogeneous bulk energy momentum tensor $T^{(B)~N}_{~~~M}=-\Lambda \delta_M^N$ but we do not impose
any constraints on the four-brane tension: 
\begin{equation}
T^{(br)~N}_{~~~M}=-\delta(\rho)\left(\begin{array}{ccc}V_0~
\delta_{\mu}^{\nu}&~&~\\~&0&~\\~&~&V_{\theta}\end{array}\right)
\end{equation}
The Einstein equations for $\sigma(\rho)$ and $\gamma(\rho)$ in this
case  give the following solutions for the warp factors:
\begin{equation}
\sigma(\rho)=\frac{\cosh^{4/5}\left[{5 \over 4}k (\rho-\rho_0) \right]}{\cosh^{4/5}\left[{5 \over 4}k\rho_0 \right]} ~~~,~~~\gamma(\rho)=R^2\frac{\cosh^{6/5}\left[{5 \over 4}k \rho_0 \right]}{\sinh^2\left[{5 \over 4}k \rho_0 \right]}\frac{\sinh^2\left[{5 \over 4}k  (\rho-\rho_0) \right]}{\cosh^{6/5}\left[{5 \over 4}k  (\rho-\rho_0) \right]}
\end{equation}
with $k^2=-{\Lambda \over 10M^4}$, where we have normalized
$\sigma(0)=1$ and $\gamma(0)=R^2$ (see fig. \ref{bigrcon}). From the above relations it is obvious  that both $\sigma(\rho)$ and
$\gamma(\rho)$ have a bounce form. However, we note that
$\gamma(\rho_{0})=0$, that is,  $\gamma(\rho)$ is vanishing at the
minimum of the warp factor $\sigma(\rho)$. This is a general
characteristic of the solutions even when the branes are non-flat. From eq.(\ref{tt}) and eq.(\ref{rr}) we can easily show that
$\gamma(\rho)=C\frac{(\sigma'(\rho))^2}{\sigma(\rho)}$ (where C is an
integration constant) which implies
that whenever we have a bounce in the warp factor the function $\gamma(\rho)$
will develop a zero.

In order to examine the nature of this singularity we study the form of the metric at the vicinity of the point
$\rho=\rho_{0}$ along the lines of \cite{Deser:1984tn,Deser:1984dr}. Taking in account that in this limit we have 
$\sigma(\rho) \rightarrow b^2$ and $\gamma(\rho) \rightarrow \beta^2 (\rho-\rho_{0})^2$
the metric becomes:
\begin{eqnarray}
ds^2=&b^2~\eta_{\mu \nu}dx^{\mu}dx^{\nu}+d\rho^2 + \beta^2
(\rho-\rho_{0})^2 d\theta^2\\
&{\rm where}~~~b^2=\cosh^{-4/5}\left[{5 \over 4}k\rho_0 \right]\nonumber\\&{\rm and}~~~\beta^2 \equiv \frac{25 k^2 R^2}{16}
\frac{\cosh^{6/5}(\frac{5}{4}k\rho_{0})}{\sinh^{2}(\frac{5}{4}k\rho_{0})}\nonumber
\end{eqnarray}

From the form of the metric, it is clear  that for general values of
the $\beta$ parameter there will be a conical singularity with a
corresponding 
deficit angle $\delta=2\pi(1-\beta)$. The existence of this conifold singularity is
connected to the presence of a 3-brane at
$\rho=\rho_{0}$. In order
to find its tension one has to carefully examine the Einstein tensor
at the vicinity of the singularity. For this reason we write the metric for 
the internal manifold in the conformally flat form:
\begin{equation}
ds^2=b^2\eta_{\mu \nu}dx^{\mu}dx^{\nu}+f(r)(dr^2 + r^2 d\theta^2)
\end{equation}
with $f(r)=r^{2(\beta-1)}$ and $\rho-\rho_{0}=\beta^{-1}r^{\beta}$. In these coordinates it is easy to see how the three brane appears at
the conifold point. The Einstein tensor can be calculated for $\rho
\rightarrow \rho_{0}$:
\begin{equation}
R_{MN}-\frac{1}{2}G_{MN}R=\left(\begin{array}{ccc}\frac{\nabla^{2}\log(f(r))}{2f(r)}~b^2~
\eta_{\mu \nu}&~&~\\~&0&~\\~&~&0\end{array}\right)
\end{equation}
where $\nabla^{2}$ is the flat two dimensional Laplacian. Now, given that $\nabla^{2}\log(r)=2 \pi \delta^{(2)}({\bf{r}})$ and by comparing with:
\begin{equation}
R_{MN}-\frac{1}{2}G_{MN}R=-\frac{V_3}{4M^{4}}\frac{\sqrt{-\hat{G}}}{\sqrt{-G}}\hat{G}_{\mu\nu}\delta^{\mu}_{M}\delta^{\nu}_{N}\delta(r)
\end{equation}
where $\delta^{(2)}({\bf{r}})={\delta(r) \over 2\pi r}$ (see Appendix
of \cite{Leblond:2001xr}), we find that the tension of the 3-brane is:
\begin{equation}
V_{3}=4(1-\beta)M^{4}={2M^4 \over \pi}\delta
\end{equation}
Thus, if there is angle deficit $\delta>0$ ($\beta<1$) the tension of
the brane is  positive, whereas if there is angle excess $\delta<0$
($\beta>1$) the tension of the brane is negative. At the critical
value $\beta=0$ there is no conical singularity at all and we have a
situation where two locally flat spaces touch each other at one point.


In the following we will concentrate on the symmetric case, that is,
the four-branes will be considered placed at symmetric points with respect to 
the minimum of the warp factor $\rho_{0}$, \textit{i.e.}
$L=2\rho_0$. For the previous solution to be consistent the tensions of the
four-branes for the symmetric configuration must satisfy: 
\begin{equation}
V_{\theta}=-{8\Lambda \over 5k}\tanh\left[{5 \over 2}k
\rho_0\right]~~~,~~~V_0={3 \over 8}V_{\theta}+{8\Lambda^2 \over
5k^2}{1 \over V_{\theta}}
\end{equation}
The brane tensions for both branes are identical. Thus, the above construction consists of two positive tension four-branes placed at the end of the compact space and an intermediate three-brane due to the conifold singularity with tension depending on the
parameters of the model. In the limit $\rho_0 \rightarrow \infty$ we
correctly obtain two identical minimal single four-brane models for the case
where $\delta \to 2\pi$ ($\beta \to 0$).

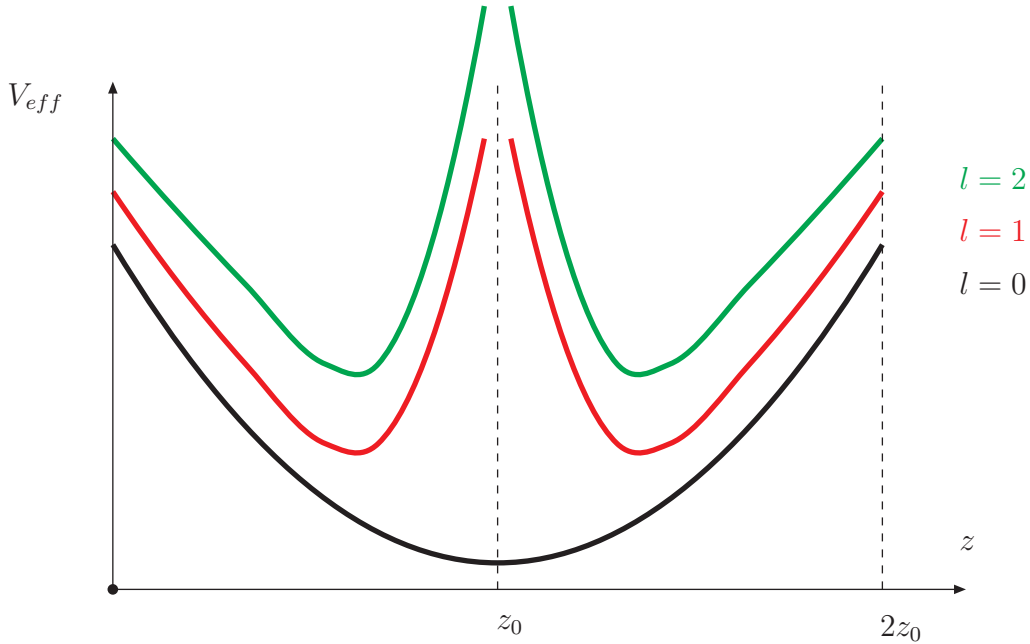
\begin{figure}[t]
\vskip20mm
\begin{center}
\begin{picture}(200,100)(0,50)

\SetWidth{2}
\SetColor{Black}
\Curve{(-70,110)(75,-10)(220,110)}
\SetColor{Red}
\Curve{(-70,130)(-20,65)(10,35)(30,35)(70,150)}
\Curve{(80,150)(120,35)(140,35)(170,65)(220,130)}
\SetColor{Green}
\Curve{(-70,150)(-20,95)(10,65)(30,65)(70,200)}
\Curve{(80,200)(120,65)(140,65)(170,95)(220,150)}
\SetColor{Black}
\Vertex(-70,-20){2}

\SetWidth{.5}
\LongArrow(-70,-20)(250,-20)
\LongArrow(-70,-20)(-70,170)
\DashLine(220,-20)(220,170){3}
\DashLine(75,-20)(75,170){3}

\SetWidth{2}
\SetColor{Black}
\Text(250,0)[lt]{$z$}
\Text(-110,160)[lb]{$V_{eff}$}
\Text(250,100)[lt]{$l=0$}
\Text(250,120)[lt]{${\Red{l=1}}$}
\Text(250,140)[lt]{${\Green{l=2}}$}
\Text(75,-30)[lt]{$z_0$}
\Text(220,-30)[lt]{$2z_0$}

\end{picture}
\end{center}

\vskip25mm
\caption{The form of the effective potential $V_{eff}$ for different angular
quantum numbers $l$ for the conifold bigravity model in the symmetric configuration. The
$\delta$-functions in the positions of the four-branes are omitted.}
\label{potbigrcon}
\end{figure}

The differential equation for the radial wavefunction $\phi$ of 
the graviton excitations is:
\begin{equation} 
\phi''+2\left({\sigma' \over \sigma}+{\gamma' \over
4\gamma}\right)\phi'+\left({m^2 \over \sigma}-{l^2  \over \gamma}\right)\phi=0
\label{diff}
\end{equation}
with normalization $\int_0^{2\rho_0} d\rho \sigma \sqrt{\gamma} \phi_m
\phi_n=\delta_{mn}$.

There is an obvious normalizable zero mode with $\phi_0=const.$ and a
tower of discrete KK states. It is easy to demonstrate that for
$\delta>0$ ($\beta<1$) this model has an ultralight state, leading to a
theory of bigravity without negative tension branes. This follows
\cite{new} considering the limit $\rho_0 \to \infty$ when the theory
describes two identical single four-brane models each of which has a
massless graviton. For finite $\rho_0$ only their symmetric combination
remains massless but their antisymmetric combination acquires an
anomalously small mass, because the difference in the modulus of the
wavefunctions between the massive and the massless states is
significant only near $\rho_0$ where the wavefunctions are
exponentially small\footnote{ There is, however, a 
mismatch of the degrees of freedom of the massive and massless
graviton, which may not be phenomenologically dangerous, but still raises the
issue of locality in the infinite four-brane separation limit.}.

We can transform this differential equation to a two
dimensional Schr\"{o}dinger-like one by the coordinate change ${dz
\over d\rho}=\sigma^{-1/2}\equiv g(z)$ and the usual wavefunction
redefinition. Then the effective potential reads:
\begin{equation}
V_{eff}={15 \over 8}\left({\de_z g \over g}\right)^2-{3 \over 4}{\de_z^2 g \over
g}-{3 \over 8}{\de_z g \de_z \gamma \over g \gamma}+{l^2 \over 2\gamma 
g^2}
\label{pot}
\end{equation}

We cannot write an explicit analytic formula for of the above
potential in the $z$-coordinates because the coordinate transformation 
is not invertible analytically. However, since the transformation is
monotonic, we can easily sketch the form of the potential (see fig.\ref{potbigrcon}) by
calculating it in the $\rho$ coordinates. From this procedure we see
that the potential for the s-wave is finite at the conical singularity 
$\rho_0$ but has a divergence for all the angular excitations with
$l\neq0$. This means that only the s-wave excitations with communicate
the two parts of the  conifold and the other excitations will be
confined in the two semicones. The spectrum of KK states in this case will be discrete. 

The construction may be readily modified to obtain a six-dimensional
analogue of the locally localized model \cite{Karch:2001ct}. This is
done by considering the asymmetric situation in which one of the fixed 
point four-branes is moved to infinity (in the asymmetric case the tensions of
the two four-branes are different). Then gravity on
the four-brane at $\rho=0$ will be mediated by only the ultralight
state since the  graviton zero mode will not be normalizable.

\subsection{The non-singular model}

\begin{figure}[t]
\vskip10mm
\begin{center}
\begin{picture}(200,100)(0,50)

\SetWidth{2}
\SetColor{Red}
\Oval(-50,100)(50,20)(0)
\SetColor{Black}
\Curve{(-50,150)(100,105)(250,150)}
\Curve{(-50,50)(100,95)(250,50)}
\Vertex(-50,100){2}
\Vertex(250,100){2}

\SetColor{Red}
\Oval(250,100)(50,20)(0)

\SetColor{Black}
\SetWidth{.5}
\LongArrow(-50,150)(-50,170)
\DashLine(-50,100)(-50,150){3}
\DashLine(-50,100)(-70,80){3}
\Line(250,100)(230,80)

\SetColor{Black}
\SetWidth{.5}
\DashLine(-50,100)(250,100){3}
\LongArrow(250,100)(300,100)

\SetWidth{3}
\SetColor{White}
\Line(-40,135)(-30,135)
\Line(-40,125)(-30,125)
\Line(-40,115)(-30,115)
\Line(-40,105)(-25,105)
\Line(-40,95)(-25,95)
\Line(-40,85)(-30,85)
\Line(-40,75)(-30,75)
\Line(-40,65)(-30,65)

\SetWidth{2}
\SetColor{Black}
\Text(300,90)[lt]{$\rho$}
\Text(80,90)[lt]{$\rho_0$}
\Text(250,110)[lb]{$L$}
\Text(-60,85)[lt]{$R$}
\Text(-70,160)[lb]{$\sigma$}

\end{picture}
\end{center}

\vskip5mm

\vskip15mm
\begin{center}
\begin{picture}(200,100)(0,50)

\SetWidth{2}
\SetColor{Red}
\Oval(-50,100)(50,20)(0)
\SetColor{Black}
\Curve{(-50,150)(40,170)(100,200)(160,170)(250,150)}
\Curve{(-50,50)(40,30)(100,0)(160,30)(250,50)}
\Vertex(-50,100){2}
\Vertex(250,100){2}

\SetColor{Red}
\Oval(250,100)(50,20)(0)

\SetColor{Black}
\SetWidth{.5}
\LongArrow(-50,150)(-50,170)
\DashLine(-50,100)(-50,150){3}
\DashLine(-50,100)(-70,80){3}
\Line(250,100)(230,80)

\SetColor{Black}
\SetWidth{.5}
\DashLine(-50,100)(250,100){3}
\LongArrow(250,100)(300,100)

\SetWidth{3}
\SetColor{White}
\Line(-40,135)(-30,135)
\Line(-40,125)(-30,125)
\Line(-40,115)(-30,115)
\Line(-40,105)(-25,105)
\Line(-40,95)(-25,95)
\Line(-40,85)(-30,85)
\Line(-40,75)(-30,75)
\Line(-40,65)(-30,65)

\SetWidth{2}
\SetColor{Black}
\Text(300,90)[lt]{$\rho$}
\Text(80,90)[lt]{$\rho_0$}
\Text(250,110)[lb]{$L$}
\Text(-60,85)[lt]{$R$}
\Text(-70,160)[lb]{$\gamma$}

\SetColor{Black}
\SetWidth{.5}
\DashLine(100,-20)(100,350){2}

\end{picture}
\end{center}

\vskip20mm

\caption{The non-singular bigravity model warp factors $\sigma(\rho)$ and
$\gamma(\rho)$ for the symmetric configuration where $L=2\rho_0$. The
$\sigma(\rho)$ warp factor has exactly the same form as in the
conifold model. The $\gamma(\rho)$ warp factor however now has an
inverse bounce form.  In this case $\gamma(\rho)$ does not vanish anywhere and thus the model is free of singularities.}
\label{bigrnosing}
\end{figure}
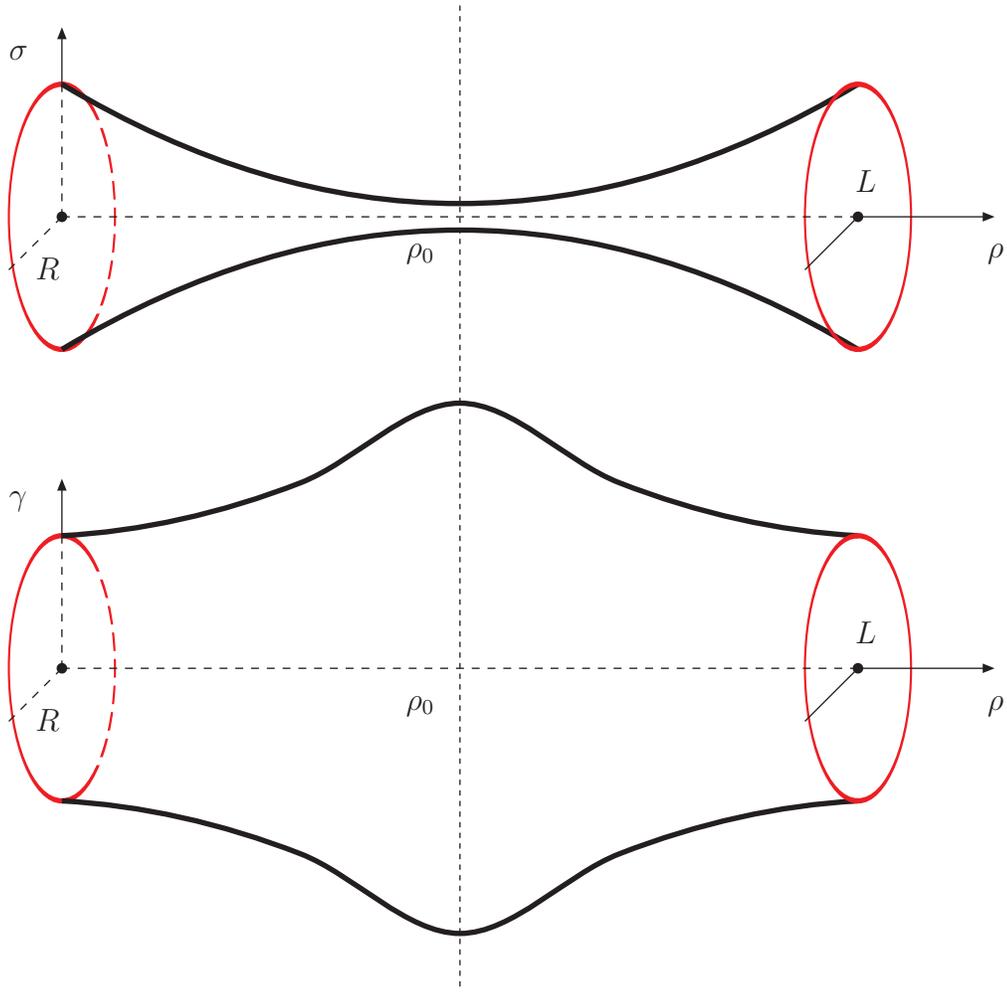

The appearance of a conifold singularity can be avoided if  instead 
of using the minimal single four-brane solution to build the double four-brane
solution, one uses the generalized single four-brane model. By
allowing for an inhomogeneous bulk
cosmological constant one can arrange that the function $\gamma(\rho)$ 
does not develop a zero. In this case the $(\theta,\theta)$ and
$(\rho,\rho)$ part of Einstein equations in the bulk, (\ref{tt}) and (\ref{rr}), have solutions: 
\begin{equation}
\sigma(\rho)=\frac{\cosh^{4/5}\left[{5 \over 4}k (\rho-\rho_0) \right]}{\cosh^{4/5}\left[{5 \over 4}k\rho_0 \right]} ~~~,~~~\gamma(\rho)=A\frac{\sinh^{\alpha}\left[{5 \over 4}k  (\rho-\rho_0) \right]}{\cosh^{6/5}\left[{5 \over 4}k  (\rho-\rho_0) \right]}
\end{equation}
where $A$ is a constant and $\alpha \equiv \frac{2
\Lambda_{\rho}}{\Lambda_{\theta}}$.  The $(\mu,\nu)$ component of the Einstein equations (\ref{mn}), however, restricts
the possible values of $\alpha$ to $\alpha=2$ or  $\alpha=0$. The case 
of $\alpha=2$ corresponds to the previous conifold model. For the 
case $\alpha=0$, which corresponds  to the choice $\Lambda_{\rho}=0$,
the solution is:  
\begin{equation}
\sigma(\rho)=\frac{\cosh^{4/5}\left[{5 \over 4}k (\rho-\rho_0) \right]}{\cosh^{4/5}\left[{5 \over 4}k\rho_0 \right]}~~~,~~~\gamma(\rho)=R^2\frac{\cosh^{6/5}\left[{5 \over 4}k \rho_0 \right]}{\cosh^{6/5}\left[{5 \over 4}k  (\rho-\rho_0) \right]}
\end{equation}
with $k^2=-{\Lambda_{\theta} \over 10M^4}$, where we have normalized
$\sigma(0)=1$ and $\gamma(0)=R^2$ (see fig. \ref{bigrnosing}). From the above the absence of any singularity is obvious since
$\gamma(\rho)$ does not vanish at any finite value of $\rho$.

The four branes will appear as $\delta$-function singularities at the
points where we will cut the space in the $\rho$
direction. In the following we will concentrate on the symmetric case, that is,
the branes will be considered placed in symmetric points in respect to 
the extremum of the warp factor $\rho_{0}$, \textit{i.e.} $L=2\rho_0$.
In order for the above configuration to be realized, the components of the four-brane tensions and
the bulk cosmological constant must be tuned to give: 
\begin{equation}
V_{\theta}=-{8\Lambda \over 5k}\tanh\left[{5 \over 2}k
\rho_0\right]~~~,~~~V_0={3 \over 8}V_{\theta}~~~,~~~\Lambda_0={3 \over 8}\Lambda_{\theta}
\end{equation}
The brane tensions for both branes are identical. Thus, we have constructed a compact model with two positive tension
flat branes with a bounce in the warp factors. Note the extra
condition relating the different components of the bulk cosmological
constant; this inhomogeneity in the bulk cosmological constant
 is what allows us to have a non-singular solution. 
In the limit $\rho_0 \rightarrow \infty$ we  obtain two identical
generalized single four-brane $\alpha=0$ models.

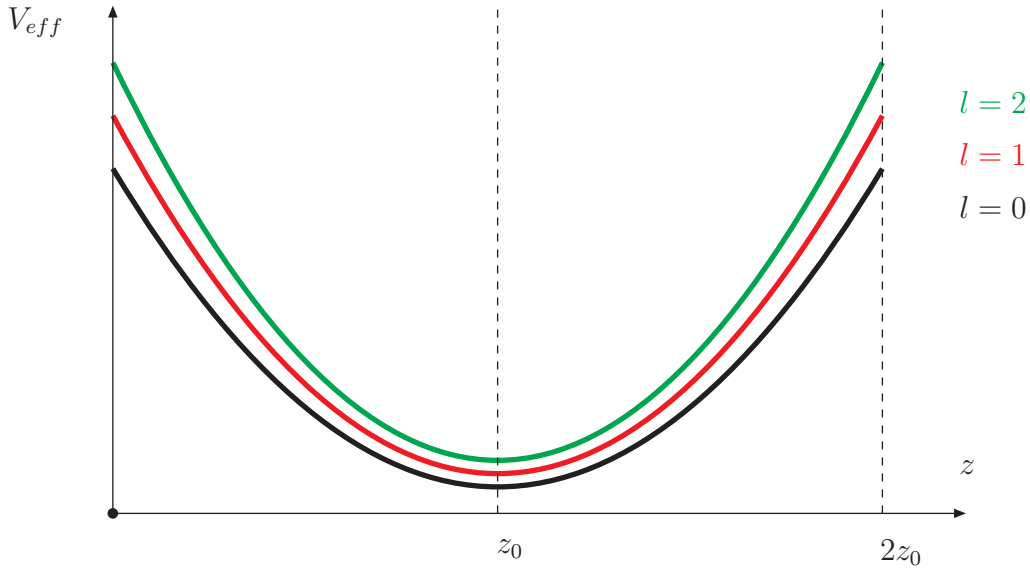
\begin{figure}[t]
\vskip20mm
\begin{center}
\begin{picture}(200,100)(0,50)

\SetWidth{2}
\SetColor{Black}
\Curve{(-70,110)(75,-10)(220,110)}
\SetColor{Red}
\Curve{(-70,130)(75,-5)(220,130)}
\SetColor{Green}
\Curve{(-70,150)(75,0)(220,150)}
\SetColor{Black}
\Vertex(-70,-20){2}

\SetWidth{.5}
\LongArrow(-70,-20)(250,-20)
\LongArrow(-70,-20)(-70,170)
\DashLine(220,-20)(220,170){3}
\DashLine(75,-20)(75,170){3}

\SetWidth{2}
\SetColor{Black}
\Text(250,0)[lt]{$z$}
\Text(-110,160)[lb]{$V_{eff}$}
\Text(250,100)[lt]{$l=0$}
\Text(250,120)[lt]{${\Red{l=1}}$}
\Text(250,140)[lt]{${\Green{l=2}}$}
\Text(75,-30)[lt]{$z_0$}
\Text(220,-30)[lt]{$2z_0$}

\end{picture}
\end{center}

\vskip25mm
\caption{The form of the effective potential $V_{eff}$ for different angular
quantum numbers $l$ for the non-singular bigravity model in the symmetric configuration. The
$\delta$-functions in the positions of the four-branes are omitted.}
\label{potbigrnosing}
\end{figure}

 The differential equation for the radial wavefunction $\phi$ of 
the graviton excitations is given by (\ref{diff}) for the relevant
$\sigma(\rho)$ and $\gamma(\rho)$  functions.
Although it is difficult to find an  analytical solution 
of the corresponding differential equation for the KK states,
the existence of a light special KK is assured from the locality
arguments discussed above \cite{new}. Since at the infinite brane
separation limit we recover two identical generalized single four-brane
models (for $\alpha=0$) where each of them supports a massless zero mode, at a finite
separation configuration the one of the zero modes will become the
 special light KK state.

By making an appropriate change of coordinates and a redefinition of
the wavefunction we can bring the previous differential equation in a
two dimensional Schr\"{o}dinger form with the effective potential (\ref{pot}) that is 
plotted in fig.\ref{potbigrnosing}. Note that in this case the angular
excitations feel the whole region
between the two positive branes and thus communicate physics between them.
The spectrum of KK states in this case will again be discrete.

\section{Multigravity in six dimensions}
  
In this section we will show how we can obtain the analogues of the
GRS model of quasi-localized gravity and  of the crystal
universe in six dimensions, in both cases  without the need of introducing moving
negative tension branes.

\begin{figure}[t]
\vskip10mm
\begin{center}
\begin{picture}(200,100)(0,50)

\SetWidth{2}
\SetColor{Red}
\Oval(-50,100)(50,20)(0)
\SetColor{Black}
\Curve{(-50,150)(100,105)(250,150)}
\Curve{(-50,50)(100,95)(250,50)}
\Vertex(-50,100){2}

\SetColor{Black}
\SetWidth{.5}
\LongArrow(-50,150)(-50,170)
\DashLine(-50,100)(-50,150){3}
\DashLine(-50,100)(-70,80){3}
\Line(250,100)(230,80)

\CBox(151,0)(250,200){White}{White}
\SetColor{Black}
\SetWidth{.5}
\DashLine(-50,100)(230,100){3}
\LongArrow(230,100)(232,100)

\SetWidth{2}
\Line(150,110)(220,110)
\Line(150,90)(220,90)

\SetWidth{3}
\SetColor{White}
\Line(-40,135)(-30,135)
\Line(-40,125)(-30,125)
\Line(-40,115)(-30,115)
\Line(-40,105)(-25,105)
\Line(-40,95)(-25,95)
\Line(-40,85)(-30,85)
\Line(-40,75)(-30,75)
\Line(-40,65)(-30,65)

\SetWidth{2}
\SetColor{Black}
\Text(240,90)[lt]{$\rho$}
\Text(80,90)[lt]{$\rho_0$}
\Text(150,80)[lt]{$\rho_b$}
\Text(-60,85)[lt]{$R$}
\Text(-70,160)[lb]{$\sigma$}

\SetWidth{2}
\SetColor{Red}
\Oval(150,100)(10,3)(0)

\SetWidth{3}
\SetColor{White}
\Line(165,105)(150,105)
\Line(165,95)(150,95)

\end{picture}
\end{center}

\vskip5mm

\vskip5mm
\begin{center}
\begin{picture}(200,100)(0,50)

\SetWidth{2}
\SetColor{Red}
\Oval(-50,100)(50,20)(0)
\SetColor{Black}
\Curve{(-50,150)(40,170)(100,200)(160,170)(250,150)}
\Curve{(-50,50)(40,30)(100,0)(160,30)(250,50)}
\Vertex(-50,100){2}

\SetColor{Black}
\SetWidth{.5}
\LongArrow(-50,150)(-50,170)
\DashLine(-50,100)(-50,150){3}
\DashLine(-50,100)(-70,80){3}
\Line(250,100)(230,80)

\CBox(151,0)(250,200){White}{White}
\SetColor{Black}
\SetWidth{.5}
\DashLine(-50,100)(230,100){3}
\LongArrow(230,100)(232,100)

\SetWidth{2}
\SetColor{Black}
\Line(150,175)(220,175)
\Line(150,25)(220,25)

\SetWidth{3}
\SetColor{White}
\Line(-40,135)(-30,135)
\Line(-40,125)(-30,125)
\Line(-40,115)(-30,115)
\Line(-40,105)(-25,105)
\Line(-40,95)(-25,95)
\Line(-40,85)(-30,85)
\Line(-40,75)(-30,75)
\Line(-40,65)(-30,65)

\SetWidth{2}
\SetColor{Black}
\Text(240,90)[lt]{$\rho$}
\Text(80,90)[lt]{$\rho_0$}
\Text(150,90)[lt]{$\rho_b$}
\Text(-60,85)[lt]{$R$}
\Text(-70,160)[lb]{$\gamma$}

\SetColor{Black}
\SetWidth{.5}
\DashLine(100,-10)(100,320){2}

\SetWidth{2}
\SetColor{Red}
\Oval(150,100)(75,20)(0)

\SetWidth{3}
\SetColor{White}
\Line(170,165)(155,165)
\Line(170,155)(160,155)
\Line(170,145)(160,145)
\Line(170,135)(160,135)
\Line(170,125)(160,125)
\Line(175,115)(160,115)
\Line(175,105)(160,105)
\Line(175,95)(160,95)
\Line(175,85)(160,85)
\Line(170,75)(160,75)
\Line(170,65)(160,65)
\Line(170,55)(160,55)
\Line(170,45)(160,45)
\Line(170,35)(155,35)

\end{picture}
\end{center}

\vskip15mm
\caption{The $\alpha=0$ quasi-localized model warp factors $\sigma(\rho)$ and
$\gamma(\rho)$. The bulk region for $\rho>\rho_b$ is flat.}
\label{quasi1}
\end{figure}
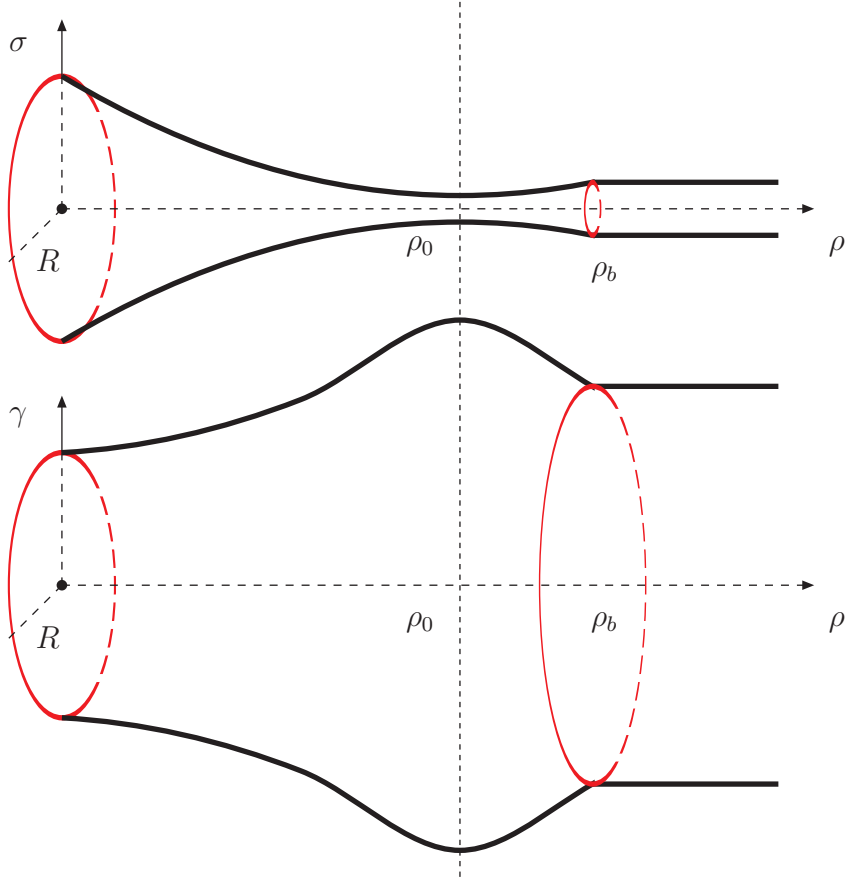

\subsection{Quasi-localized gravity}

As in the bigravity models we have two possible ways of realizing a
quasi-localized scenario in six dimensions. The first model can be built
from the $\alpha=0$ bigravity model, by cutting the space at a point
$\rho_b$ on the right 
of the position of the bounce and sewing it to flat space on the right (see
fig.\ref{quasi1}). In that case we will have a system of two positive tension
branes and by appropriately tuning the different components of the
brane tension of the second brane we can find solutions where the warp factor is constant
in the flat bulk region. The fine tunings which achieve this are:
\begin{equation}
V^{(2)}_{\theta}=-{4\Lambda \over 5k}\tanh\left[{5 \over 2}k (\rho_b-\rho_0)\right]~~~,~~~V^{(2)}_0={3 \over 8}V^{(2)}_{\theta}
\end{equation}

From the above formulas it is obvious that if we paste the flat space
at the position of the bounce, \textit{i.e} $\rho_b=\rho_0$ then the
second brane is tensionless. If we relax these fine tunings we can find 
solutions where the warp factors in the flat bulk region are non
trivial, but we will not discuss this possibility here.

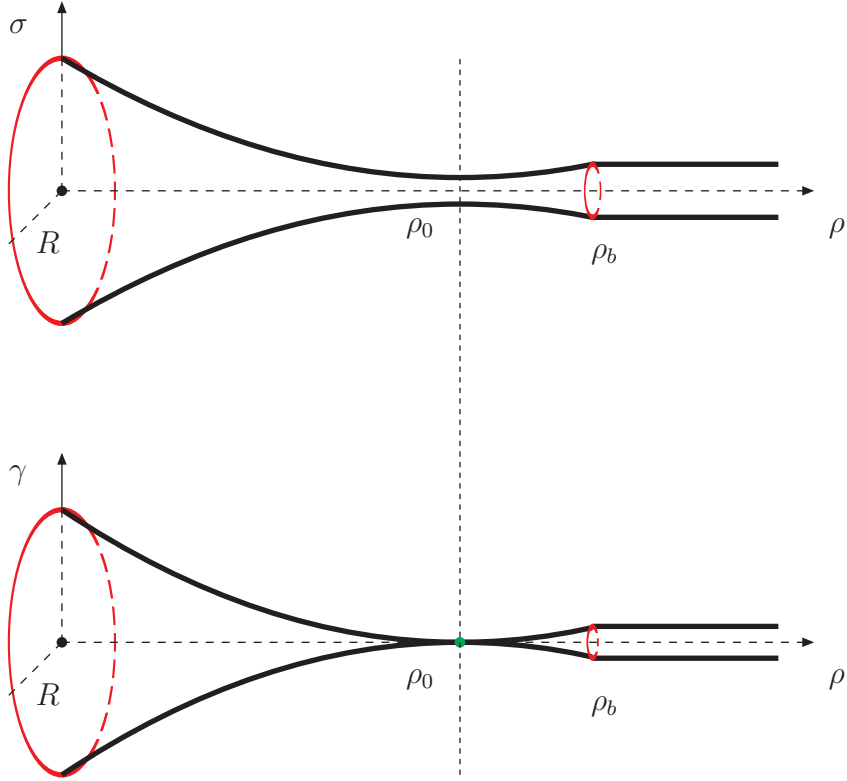
\begin{figure}[t]
\vskip10mm
\begin{center}
\begin{picture}(200,100)(0,50)

\SetWidth{2}
\SetColor{Red}
\Oval(-50,100)(50,20)(0)
\SetColor{Black}
\Curve{(-50,150)(100,105)(250,150)}
\Curve{(-50,50)(100,95)(250,50)}
\Vertex(-50,100){2}

\SetColor{Black}
\SetWidth{.5}
\LongArrow(-50,150)(-50,170)
\DashLine(-50,100)(-50,150){3}
\DashLine(-50,100)(-70,80){3}
\Line(250,100)(230,80)

\CBox(151,0)(250,200){White}{White}
\SetColor{Black}
\SetWidth{.5}
\DashLine(-50,100)(230,100){3}
\LongArrow(230,100)(232,100)

\SetWidth{2}
\Line(150,110)(220,110)
\Line(150,90)(220,90)

\SetWidth{3}
\SetColor{White}
\Line(-40,135)(-30,135)
\Line(-40,125)(-30,125)
\Line(-40,115)(-30,115)
\Line(-40,105)(-25,105)
\Line(-40,95)(-25,95)
\Line(-40,85)(-30,85)
\Line(-40,75)(-30,75)
\Line(-40,65)(-30,65)

\SetWidth{2}
\SetColor{Black}
\Text(240,90)[lt]{$\rho$}
\Text(80,90)[lt]{$\rho_0$}
\Text(150,80)[lt]{$\rho_b$}
\Text(-60,85)[lt]{$R$}
\Text(-70,160)[lb]{$\sigma$}

\SetWidth{2}
\SetColor{Red}
\Oval(150,100)(10,3)(0)

\SetWidth{3}
\SetColor{White}
\Line(165,105)(150,105)
\Line(165,95)(150,95)

\end{picture}
\end{center}

\vskip5mm

\vskip15mm
\begin{center}
\begin{picture}(200,100)(0,50)

\SetWidth{2}
\SetColor{Red}
\Oval(-50,100)(50,20)(0)
\SetColor{Black}
\Curve{(-50,150)(100,100)(250,150)}
\Curve{(-50,50)(100,100)(250,50)}
\Vertex(-50,100){2}

\CBox(151,0)(250,200){White}{White}
\SetColor{Black}
\SetWidth{.5}
\LongArrow(-50,150)(-50,170)
\DashLine(-50,100)(-50,150){3}
\DashLine(-50,100)(-70,80){3}
\Line(250,100)(230,80)

\CBox(151,0)(250,200){White}{White}
\SetColor{Black}
\SetWidth{.5}
\DashLine(-50,100)(230,100){3}
\LongArrow(230,100)(232,100)

\SetWidth{2}
\Line(150,106)(220,106)
\Line(150,94)(220,94)

\SetWidth{3}
\SetColor{White}
\Line(-40,135)(-30,135)
\Line(-40,125)(-30,125)
\Line(-40,115)(-30,115)
\Line(-40,105)(-25,105)
\Line(-40,95)(-25,95)
\Line(-40,85)(-30,85)
\Line(-40,75)(-30,75)
\Line(-40,65)(-30,65)

\SetWidth{2}
\SetColor{Black}
\Text(240,90)[lt]{$\rho$}
\Text(80,90)[lt]{$\rho_0$}
\Text(150,80)[lt]{$\rho_b$}
\Text(-60,85)[lt]{$R$}
\Text(-70,160)[lb]{$\gamma$}

\SetWidth{2}
\SetColor{Green}
\Vertex(100,100){2}

\SetWidth{2}
\SetColor{Red}
\Oval(150,100)(6,2)(0)

\SetWidth{2}
\SetColor{White}
\Line(165,102.5)(150,102.5)
\Line(165,97.5)(150,97.5)

\SetColor{Black}
\SetWidth{.5}
\DashLine(100,50)(100,320){2}

\end{picture}
\end{center}

\caption{The $\alpha=2$ quasi-localized model warp factors $\sigma(\rho)$ and
$\gamma(\rho)$. The bulk region for $\rho>\rho_b$ is flat.}
\label{quasi2}
\end{figure}

The above construction will obviously have no normalizable zero mode
due to the infinite volume of the system and the KK spectrum will be
continuous. However, as in the five dimensional case we expect that
the low part of the KK tower will have a ``resonance''-like coupling
to matter on the $''+''$  brane at $\rho=0$. This will mediate normal 
gravity at intermediate distances but will change the nature of the
gravitational law at ultralarge scales.

\begin{figure}[t]
\vskip10mm
\begin{center}
\begin{picture}(200,100)(0,50)

\SetWidth{2}
\SetColor{Red}
\Oval(-50,100)(50,20)(0)
\SetColor{Black}
\Curve{(-50,150)(25,105)(100,150)}
\Curve{(100,150)(175,105)(250,150)}
\Curve{(250,150)(325,105)(400,150)}
\Curve{(-200,150)(-125,105)(-50,150)}

\Curve{(-200,50)(-125,95)(-50,50)}
\Curve{(250,50)(325,95)(400,50)}
\Curve{(-50,50)(25,95)(100,50)}
\Curve{(100,50)(175,95)(250,50)}
\Vertex(100,100){2}

\SetColor{Red}
\Oval(250,100)(50,20)(0)
\Oval(100,100)(50,20)(0)

\CBox(-200,0)(-100,200){White}{White}
\CBox(300,0)(400,200){White}{White}
\SetColor{Black}
\SetWidth{.5}
\LongArrow(100,150)(100,170)
\DashLine(100,100)(100,150){3}
\DashLine(-50,100)(-70,80){3}

\SetColor{Black}
\SetWidth{.5}
\DashLine(-100,100)(300,100){3}
\LongArrow(300,100)(301,100)

\SetWidth{3}
\SetColor{White}
\Line(-40,125)(-30,125)
\Line(-40,115)(-30,115)
\Line(-40,105)(-25,105)
\Line(-40,95)(-25,95)
\Line(-40,85)(-30,85)
\Line(-40,75)(-30,75)

\Line(110,125)(120,125)
\Line(110,115)(120,115)
\Line(110,105)(125,105)
\Line(110,95)(125,95)
\Line(110,85)(120,85)
\Line(110,75)(120,75)

\Line(260,125)(270,125)
\Line(260,115)(270,115)
\Line(260,105)(275,105)
\Line(260,95)(275,95)
\Line(260,85)(270,85)
\Line(260,75)(270,75)

\SetWidth{2}
\SetColor{Black}
\Text(300,90)[lt]{$\rho$}
\Text(-60,85)[lt]{$R$}
\Text(80,160)[lb]{$\sigma$}
\Text(175,90)[lt]{$\rho_0$}
\Text(15,90)[lt]{$-\rho_0$}
\Text(310,110)[lt]{${\Huge{\bf{\cdots}}}$}
\Text(-110,110)[rt]{${\Huge{\bf{\cdots}}}$}

\end{picture}
\end{center}

\vskip5mm

\vskip15mm
\begin{center}
\begin{picture}(200,100)(0,50)

\SetWidth{2}
\SetColor{Red}
\Oval(-50,100)(50,20)(0)
\SetColor{Black}
\Curve{(-50,150)(-10,158)(25,175)(60,158)(100,150)}
\Curve{(100,150)(140,158)(175,175)(210,158)(250,150)}
\Curve{(250,150)(290,158)(325,175)(360,158)(400,150)}
\Curve{(-200,150)(-160,158)(-125,175)(-90,158)(-50,150)}

\Curve{(-200,50)(-160,42)(-125,25)(-90,42)(-50,50)}
\Curve{(250,50)(290,42)(325,25)(360,42)(400,50)}
\Curve{(-50,50)(-10,42)(25,25)(60,42)(100,50)}
\Curve{(100,50)(140,42)(175,25)(210,42)(250,50)}
\Vertex(100,100){2}

\SetColor{Red}
\Oval(250,100)(50,20)(0)
\Oval(100,100)(50,20)(0)

\CBox(-200,0)(-100,200){White}{White}
\CBox(300,0)(400,200){White}{White}
\SetColor{Black}
\SetWidth{.5}
\LongArrow(100,150)(100,170)
\DashLine(100,100)(100,150){3}
\DashLine(-50,100)(-70,80){3}

\SetColor{Black}
\SetWidth{.5}
\DashLine(-100,100)(300,100){3}
\LongArrow(300,100)(301,100)

\SetWidth{3}
\SetColor{White}
\Line(-40,135)(-30,135)
\Line(-40,125)(-30,125)
\Line(-40,115)(-30,115)
\Line(-40,105)(-25,105)
\Line(-40,95)(-25,95)
\Line(-40,85)(-30,85)
\Line(-40,75)(-30,75)
\Line(-40,65)(-30,65)

\Line(110,135)(120,135)
\Line(110,125)(120,125)
\Line(110,115)(120,115)
\Line(110,105)(125,105)
\Line(110,95)(125,95)
\Line(110,85)(120,85)
\Line(110,75)(120,75)
\Line(110,65)(120,65)

\Line(260,135)(270,135)
\Line(260,125)(270,125)
\Line(260,115)(270,115)
\Line(260,105)(275,105)
\Line(260,95)(275,95)
\Line(260,85)(270,85)
\Line(260,75)(270,75)
\Line(260,65)(270,65)

\SetWidth{2}
\SetColor{Black}
\Text(300,90)[lt]{$\rho$}
\Text(-60,85)[lt]{$R$}
\Text(80,160)[lb]{$\gamma$}
\Text(175,90)[lt]{$\rho_0$}
\Text(15,90)[lt]{$-\rho_0$}
\Text(310,110)[lt]{${\Huge{\bf{\cdots}}}$}
\Text(-110,110)[rt]{${\Huge{\bf{\cdots}}}$}

\end{picture}
\end{center}

\vskip5mm
\caption{The $\alpha=0$ crystalline model warp factors $\sigma(\rho)$ and
$\gamma(\rho)$.}
\label{crystal1}
\end{figure}
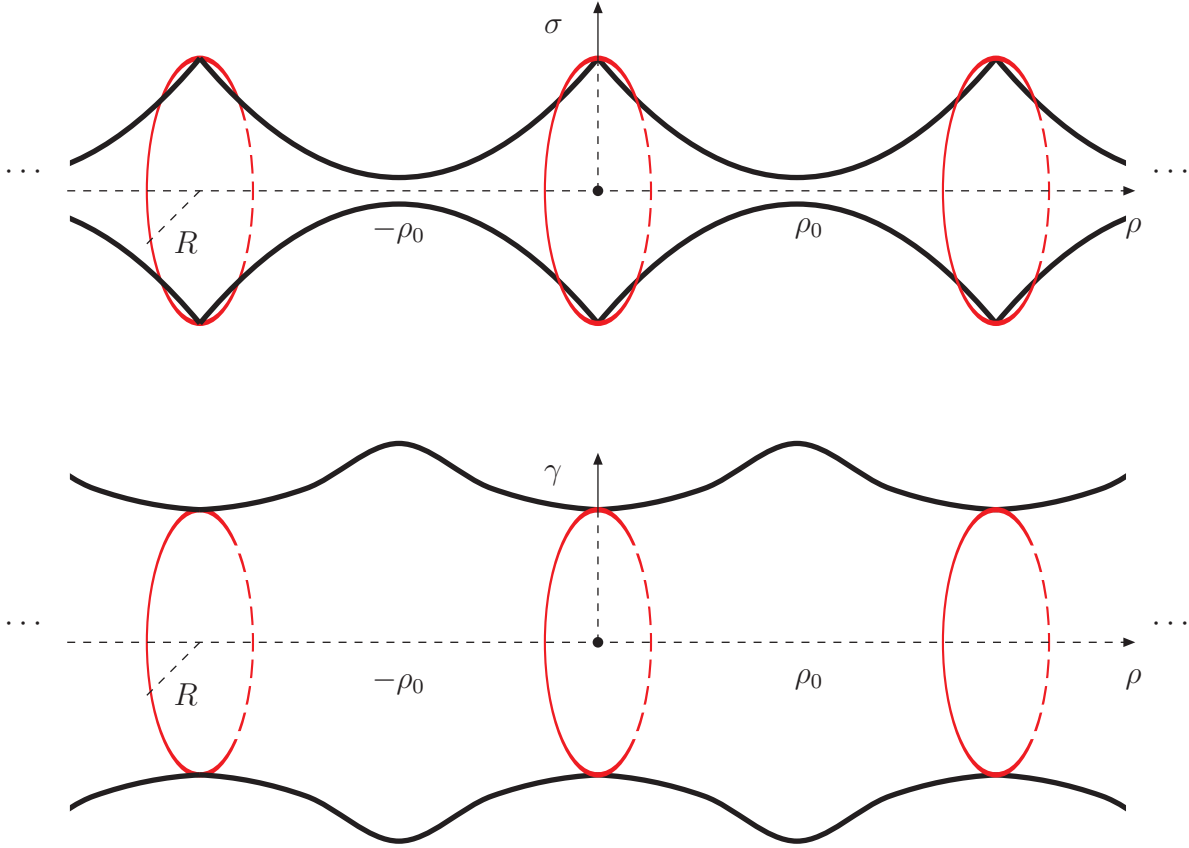

The second model of quasi-localized gravity can be built in a similar
fashion from the $\alpha=2$ bigravity model (see fig.\ref{quasi2}). Again we obtain a system
of two positive tension branes which will have a constant warp factor
in the flat bulk region when the following fine tunings are demanded:
\begin{equation}
V^{(2)}_{\theta}=-{4\Lambda \over 5k}\tanh\left[{5 \over 2}k (\rho_b-\rho_0)\right]~~~,~~~V^{(2)}_0={3 \over 8}V^{(2)}_{\theta}+{8\Lambda^2 \over 5k^2}{1 \over V^{(2)}_{\theta}}
\end{equation}

From the above formulas it is obvious that if we paste the flat space
at the position of the bounce, \textit{i.e} $\rho_b=\rho_0$ then the
tension $V^{(2)}_{\theta}$ vanishes whereas the tension $V^{(2)}_0$
diverges.  This limit is not well behaved since the internal space
becomes singular in a whole line. At this limit obviously classical
gravity breaks down and one would expect that quantum gravity
corrections would resolve the singularity. However, one should check if these quantum gravity corrections affect the
solution in the vicinity of the four-brane at $\rho=0$, \textit{e.g.} by examining the
$R^2$ correction to the Einstein action\footnote{We would like to thank Tony
Gherghetta for this comment.}.

The above construction will again have no normalizable zero mode and
the KK spectrum will be continuous. A ``resonance''-like coupling of
the KK states to matter on the $''+''$  brane at $\rho=0$ is again
expected which will give normal  four dimensional gravity at
intermediate distances but will change the nature of the gravitational law at ultralarge scales.

\begin{figure}[t]
\vskip10mm
\begin{center}
\begin{picture}(200,100)(0,50)

\SetWidth{2}
\SetColor{Red}
\Oval(-50,100)(50,20)(0)
\SetColor{Black}
\Curve{(-50,150)(25,105)(100,150)}
\Curve{(100,150)(175,105)(250,150)}
\Curve{(250,150)(325,105)(400,150)}
\Curve{(-200,150)(-125,105)(-50,150)}

\Curve{(-200,50)(-125,95)(-50,50)}
\Curve{(250,50)(325,95)(400,50)}
\Curve{(-50,50)(25,95)(100,50)}
\Curve{(100,50)(175,95)(250,50)}
\Vertex(100,100){2}

\SetColor{Red}
\Oval(250,100)(50,20)(0)
\Oval(100,100)(50,20)(0)

\CBox(-200,0)(-100,200){White}{White}
\CBox(300,0)(400,200){White}{White}
\SetColor{Black}
\SetWidth{.5}
\LongArrow(100,150)(100,170)
\DashLine(100,100)(100,150){3}
\DashLine(-50,100)(-70,80){3}

\SetColor{Black}
\SetWidth{.5}
\DashLine(-100,100)(300,100){3}
\LongArrow(300,100)(301,100)

\SetWidth{3}
\SetColor{White}
\Line(-40,125)(-30,125)
\Line(-40,115)(-30,115)
\Line(-40,105)(-25,105)
\Line(-40,95)(-25,95)
\Line(-40,85)(-30,85)
\Line(-40,75)(-30,75)

\Line(110,125)(119,125)
\Line(110,115)(120,115)
\Line(110,105)(125,105)
\Line(110,95)(125,95)
\Line(110,85)(120,85)
\Line(110,75)(119,75)

\Line(260,125)(269,125)
\Line(260,115)(270,115)
\Line(260,105)(275,105)
\Line(260,95)(275,95)
\Line(260,85)(270,85)
\Line(260,75)(269,75)

\SetWidth{2}
\SetColor{Black}
\Text(300,90)[lt]{$\rho$}
\Text(-60,85)[lt]{$R$}
\Text(80,160)[lb]{$\sigma$}
\Text(175,90)[lt]{$\rho_0$}
\Text(15,90)[lt]{$-\rho_0$}
\Text(310,110)[lt]{${\Huge{\bf{\cdots}}}$}
\Text(-110,110)[rt]{${\Huge{\bf{\cdots}}}$}

\end{picture}
\end{center}

\vskip5mm

\vskip15mm
\begin{center}
\begin{picture}(200,100)(0,50)

\SetWidth{2}
\SetColor{Red}
\Oval(-50,100)(50,20)(0)
\SetColor{Black}
\Curve{(-50,150)(25,100)(100,150)}
\Curve{(100,150)(175,100)(250,150)}
\Curve{(250,150)(325,100)(400,150)}
\Curve{(-200,150)(-125,100)(-50,150)}

\Curve{(-200,50)(-125,100)(-50,50)}
\Curve{(250,50)(325,100)(400,50)}
\Curve{(-50,50)(25,100)(100,50)}
\Curve{(100,50)(175,100)(250,50)}
\Vertex(100,100){2}

\SetColor{Red}
\Oval(250,100)(50,20)(0)
\Oval(100,100)(50,20)(0)

\CBox(-200,0)(-100,200){White}{White}
\CBox(300,0)(400,200){White}{White}
\SetColor{Black}
\SetWidth{.5}
\LongArrow(100,150)(100,170)
\DashLine(100,100)(100,150){3}
\DashLine(-50,100)(-70,80){3}

\SetColor{Black}
\SetWidth{.5}
\DashLine(-100,100)(300,100){3}
\LongArrow(300,100)(301,100)

\SetWidth{3}
\SetColor{White}
\Line(-40,125)(-31,125)
\Line(-40,115)(-30,115)
\Line(-40,105)(-25,105)
\Line(-40,95)(-25,95)
\Line(-40,85)(-30,85)
\Line(-40,75)(-31,75)

\Line(110,125)(119,125)
\Line(110,115)(120,115)
\Line(110,105)(125,105)
\Line(110,95)(125,95)
\Line(110,85)(120,85)
\Line(110,75)(119,75)

\Line(260,125)(269,125)
\Line(260,115)(270,115)
\Line(260,105)(275,105)
\Line(260,95)(275,95)
\Line(260,85)(270,85)
\Line(260,75)(269,75)

\SetWidth{2}
\SetColor{Black}
\Text(300,90)[lt]{$\rho$}
\Text(-60,85)[lt]{$R$}
\Text(80,160)[lb]{$\gamma$}
\Text(175,90)[lt]{$\rho_0$}
\Text(15,90)[lt]{$-\rho_0$}
\Text(310,110)[lt]{${\Huge{\bf{\cdots}}}$}
\Text(-110,110)[rt]{${\Huge{\bf{\cdots}}}$}

\SetWidth{2}
\SetColor{Green}
\Vertex(25,100){3}
\Vertex(175,100){3}

\end{picture}
\end{center}


\caption{The $\alpha=2$ crystalline model warp factors $\sigma(\rho)$ and
$\gamma(\rho)$.}
\label{crystal2}
\end{figure}
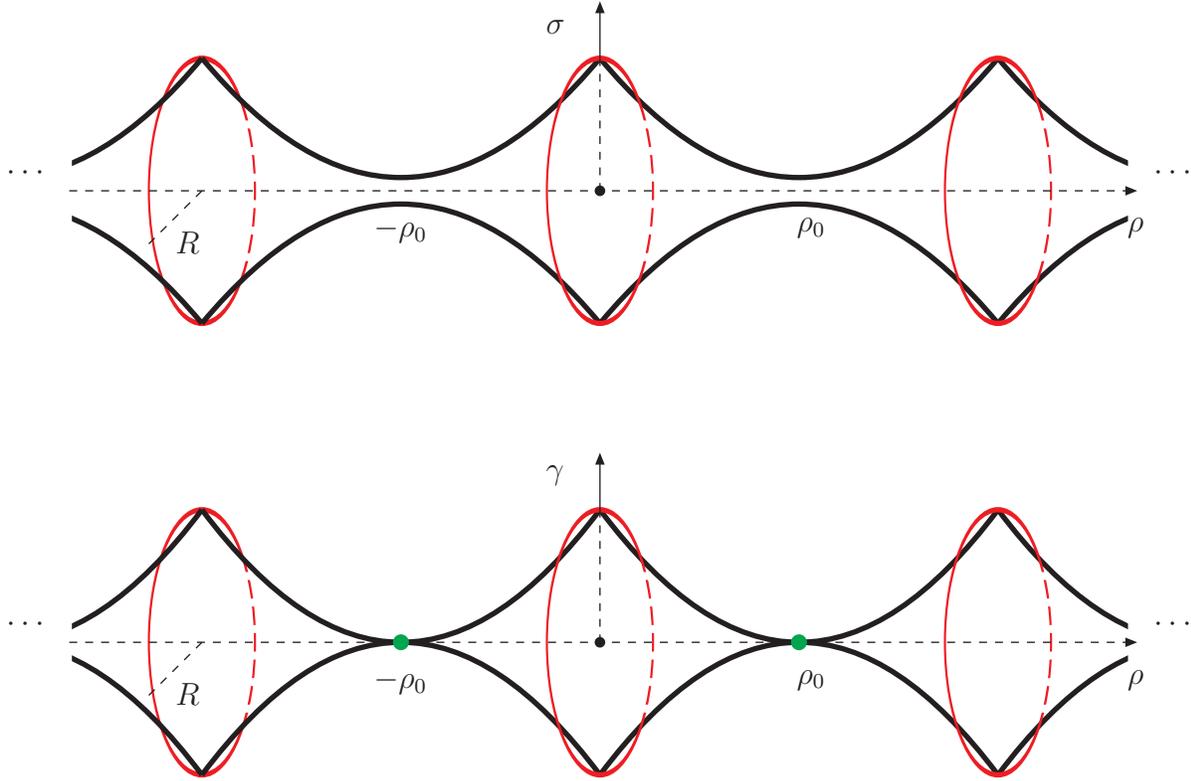

\subsection{The crystal universe model}

Another obvious generalization of the five dimensional multigravity
models is the one of the crystalline brane model. All that one has to
do is to paste an infinite array of bigravity models. We again take
this constructions in two copies. One for  $\alpha=0$ (fig.\ref{crystal1}) and one for
$\alpha=2$ (fig.\ref{crystal2}). A band structure is again expected as in the five
dimensional case and the width of the first band will be exponentially 
smaller than the one of the first forbidden zone. As far as the
phenomenology is concerned, these models will generate normal gravity
at intermediate scales as the first band will behave as an effective
zero mode, whereas at ultralarge scales will have the same change of
the gravitational law as the one of the corresponding quasi-localized models.

\section{Discussion and conclusions}   

In this paper we have constructed, for the first time, flat brane theories
which can lead to multigravity models and their associated
modifications of gravity at large distances {\textit{without}}
introducing moving negative tension branes. The constructions are made
possible by going to a six dimensional theory. In five dimensions with 
flat branes, the presence of a bounce of the warp factor was linked
to the violation of the weaker energy condition. This is not true,
however, in six dimensions as we will show in this section. 
In five dimensions with the metric:
\begin{equation}
ds^2=e^{-A(\rho)}\eta_{\mu \nu}dx^{\mu}dx^{\nu}+d\rho^2
\end{equation}
one can readily show that the weaker energy condition requires that:
\begin{equation}
A''\geq 0
\end{equation}
which is violated at the position of the moving negative tension
branes. In the case that the branes were $AdS$ one could have a bounce (without
having negative tension branes) and still satisfy the weaker energy
condition because the above relation is modified to \cite{Karch:2001ct}:
\begin{equation}
A''\geq -2H^2 e^{A}
\end{equation}
However, such models do not lead to modifications of gravity at large
distances and moreover the remnant negative cosmological constant is
in conflict with current observations.

In the six dimensional case with metric:
\begin{equation}
ds^2=e^{-A(\rho)}\eta_{\mu \nu}dx^{\mu}dx^{\nu}+d\rho^2+e^{-B(\rho)}d\theta^2
\end{equation}
from the weaker energy condition one finds two inequalities which can be cast into the following relation:
\begin{equation}
{1 \over 6} (B')^2 -{1 \over 3}B''-{1 \over 6}A'B'\leq A'' \leq -{1
\over 2} (B')^2 +B''-{3 \over 2}A'B'+ 2(A')^2
\end{equation}

These inequalities hold for the $\alpha=2$ models as long as the
three-branes sitting on the conical singularities have positive
tension. It is violated, however, in the $\alpha=0$ models everywhere
in the bulk. This shows 
that in six dimensional flat brane models, the presence of bounces of the warp factor is not
necessarily linked to whether the weaker energy condition is satisfied
or not. It is not yet clear if the violation of the weaker energy
condition in the  $\alpha=0$ models is  a sign of a possible instability.

There is an interesting feature of the conifold model that 
emerges if we consider Euclidean four dimensional space and Wick rotate 
the $\theta$ coordinate ($\theta \to it$) decompactifying it at the same time. In the case where
the deficit angle is zero, the metric in the vicinity of $\rho=\rho_0$ 
will look like:
\begin{equation}
ds^2=b^2\delta_{\mu \nu}dx^{\mu}dx^{\nu}+d\rho^2-(\rho-\rho_0)^2dt^2
\end{equation}
This is Rindler space which is deformed away of  $\rho=\rho_0$ and has sources (the four-branes) at finite proper distances from
$\rho=\rho_0$. The picture that we obtain in Kruskal coordinates for
this case is reminiscent of the one of a Schwarzschild black hole if one 
rotates the Penrose diagram at a right angle.

In summary, in this paper we discussed how we can obtain viable
multigravity models in six dimensions. The models we considered
involved flat positive tension branes (except the case when the
conifold models had angle excess). We showed how we can obtain the
double brane bigravity model, the quasi-localized model and the
crystalline model, each of which in  two types. One of them is singularity free 
while the other has conical singularities in the bulk corresponding to
three-branes. 

Several issues are still open  regarding these
constructions. Firstly, a careful treatment of the angular
excitations in the case of $\alpha=0$ should be carried out to
examine the nature of gravity on the four-branes. Furthermore, one
should examine the moduli of the system, the  radions and the dilaton,
and calculate their mass (see  \cite{Kogan:2001qx,Papazoglou:2001ed}
for the warped five dimensional case and \cite{Kanti:2001vb} for the
warped six
dimensional case). It is quite probable that the modulus
corresponding to $\rho_0$ is massive as in the five dimensional $AdS_4$
branes case and so the system is self-stabilized. Moreover, a
discussion of multigravity in a cosmological setting is yet to be
developed (for a discussion about cosmology in brane world models see for
example \cite{Binetruy:2000ut,Kanti:1999sz,Binetruy:2000hy,Csaki:2000mp,Kanti:2000nz,Kanti:2000cz,Santos:2001nt,Anchordoqui:2001qc}). Finally, the non-trivial
way that locality is preserved by these model should be studied.

\textbf{Acknowledgments:} We would like to sincerely thank Tony Gherghetta,
Panagiota Kanti, Shinsuke Kawai, Luigi Pilo and Nuno Reis for  very stimulating and informative discussions. SM's work is supported by the Hellenic State   
Scholarship Foundation (IKY) \mbox{No.    
8117781027}. AP's work is supported by the Hellenic State Scholarship   
Foundation (IKY) \mbox{No. 8017711802}. This work   is   
supported in part by the PPARC rolling grant PPA/G/O/1998/00567, by   
the EC TMR grants  HRRN-CT-2000-00148 and  HPRN-CT-2000-00152.

  \newpage

\def\theequation{A.\arabic{equation}}   
\setcounter{equation}{0}   
\vskip0.8cm   
\noindent   
\centerline{\Large \bf Appendix}   
\vskip0.4cm   
\noindent   

In this Appendix we provide the Einstein equations for the class of
models that we considered where the metric is:
\begin{equation}
ds^2=\sigma(\rho)\eta_{\mu \nu}dx^{\mu}dx^{\nu}+d\rho^2+\gamma(\rho)d\theta^2
\end{equation}
Ignoring the three-branes located on the
conical singularities, the Einstein equations are generally written as:
\begin{equation}
G_{MN}={1 \over 4M^4}(T^{(B)}_{MN}+T^{(br)}_{MN})
\end{equation}
where $T^{(B)}_{MN}$ is the bulk energy momentum tensor of the general 
form (\ref{bulkT}) and
$T^{(br)}_{MN}$ the four-brane energy momentum tensor of the form (\ref{braneT}).

In the absence of a four dimensional cosmological constant, the $(\theta,\theta)$ component of the above equation is:
\begin{equation}
2{\sigma'' \over \sigma}+{1 \over 2}\left({\sigma' \over 
\sigma}\right)^2=-{\Lambda_{\theta} \over 4M^4}-{V^i_{\theta} \over
4M^4}\delta(\rho-\rho_i)
\label{tt}
\end{equation}
The $(\rho,\rho)$ component is:
\begin{equation}
{3 \over 2}\left({\sigma' \over \sigma}\right)^2+{\sigma'\gamma' \over 
\sigma \gamma}=-{\Lambda_{\rho} \over 4M^4}
\label{rr}
\end{equation}
Finally, the $(\mu,\nu)$ component is:
\begin{equation}
{3 \over 2}{\sigma'' \over \sigma}+{3 \over 4}{\sigma'\gamma' \over
\sigma \gamma}-{1 \over 4}\left({\gamma' \over \gamma}\right)^2+{1
\over 2}{\gamma'' \over \gamma}=-{\Lambda_0 \over 4M^4}-{V^i_0 \over 4M^4}\delta(\rho-\rho_i)
\label{mn}
\end{equation}

These equations may be compared with the ones of the five
dimensional case where the metric:
\begin{equation}
ds^2=\sigma(\rho)\eta_{\mu \nu}dx^{\mu}dx^{\nu}+d\rho^2
\end{equation}
gives rise to the $(\rho,\rho)$ component:
\begin{equation}
{3 \over 2}\left({\sigma' \over \sigma}\right)^2=-{\Lambda \over 4M^3}
\label{rr5}
\end{equation}
and the $(\mu,\nu)$ component:
\begin{equation}
{3 \over 2}{\sigma'' \over \sigma}=-{\Lambda \over 4M^3}-{V^i \over 4M^3}\delta(\rho-\rho_i)
\label{mn5}
\end{equation}

The extra freedom that we have in the six dimensional case is apparent.

\end{document}